\documentclass[a4paper,11pt]{article}

\usepackage{amsfonts}
\usepackage{amsmath}
\usepackage{amssymb}
\usepackage{graphicx}
\usepackage{amscd}
\usepackage{tikz}
\usepackage{epsfig}
\usepackage{psfrag}
\usepackage{geometry}

\newtheorem{proposition}{Proposition}

\def\max{\mathop{\rm max}}

\def\Z{\mathbb{Z}}
\def\N{\mathbb{N}}

\def\P{\mathcal{{P}}}
\title{\textbf{Torus knots and generalized Schr\"oder paths}}
\date{}
\author{\textbf{Marko Sto\v si\'c$^{1,2}$ and Piotr Su\l kowski$^3$} \\ 
\\
\small{\emph{$^1$ CAMGSD, Department of Mathematics, Instituto Superior T{\'e}cnico, }}\\ 
\small{\emph{Av.  Rovisco Pais, 1049-001 Lisbon, Portugal}}\\
\small{\emph{$^2$ Mathematical Institute SANU, Knez Mihajlova 36, 11000 Beograd, Serbia}} \\
\small{\emph{$^3$ Faculty of Physics, University of Warsaw, ul. Pasteura 5, 02-093 Warsaw, Poland}}
}

\begin{document}
\maketitle

\begin{abstract}
We relate invariants of torus knots to the counts of a class of lattice paths, which we call generalized Schr{\"o}der paths. We determine generating functions of such paths, located in a region determined by a type of a torus knot under consideration, and show that they encode colored HOMFLY-PT polynomials of this knot.  The generators of uncolored HOMFLY-PT homology correspond to a basic set of such paths. Invoking the knots-quivers correspondence,  we express generating functions of such paths as quiver generating series, and also relate them to quadruply-graded knot homology.  Furthermore,  we determine corresponding A-polynomials, which provide algebraic equations and recursion relations for generating functions of generalized Schr\"oder paths.  The lattice paths of our interest explicitly enumerate BPS states associated to knots via brane constructions.
%
\end{abstract}


\newpage

\section{Introduction}

It is expected, both from mathematical and physical perspectives, that polynomial knot invariants should have a counting interpretation. From mathematical viewpoint it should arise in consequence of the structure of knot homologies, while in physics it follows from the relation between knot invariants and counting of BPS states.  In this work we find a counting interpretation of polynomial knot invariants of torus knots in terms of counting of lattice paths.  These results not only provide an explicit manifestation of a combinatorial character of knot invariants, but also lead to explicit formulae for previously unknown counting functions for various classes of lattice paths, as well as recursion relations and algebraic equations that they satisfy.

\begin{figure}[h]
  \centering
  \includegraphics[width=0.6\textwidth]{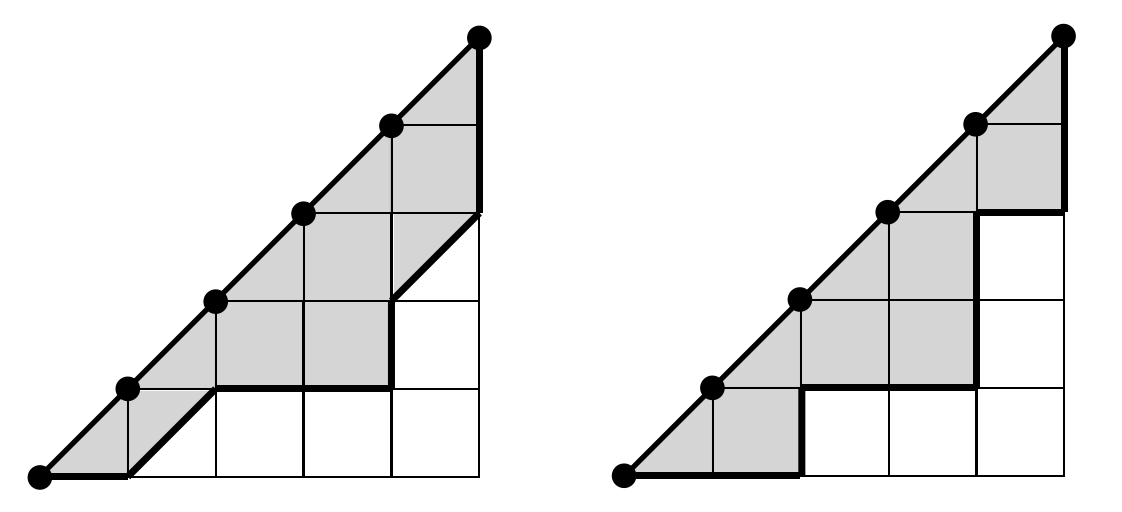}
  \caption{Schr{\"o}der paths, located below the diagonal $y=x$, encode colored HOMFLY-PT invariants of the unknot.  In general,  such paths include diagonal steps, which encode the dependence on variable $a$ (left).  Paths without diagonal steps (Dyck paths, right) encode extremal invariants (without $a$-dependence) and are enumerated by Catalan numbers. The area of the shaded region, spanned between the path and the diagonal $y=x$, captures the dependence on $q$.}  \label{fig-unknot}
\end{figure}

\begin{figure}[h]
  \centering
  \includegraphics[width=0.35\textwidth]{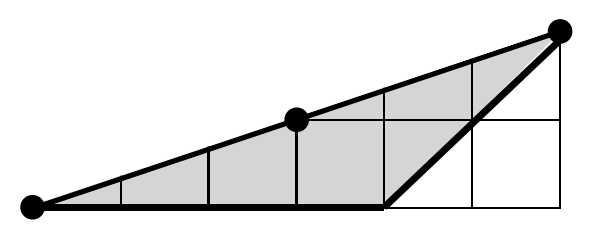}
  \caption{Generalized Schr{\"o}der paths under the line $y=\frac1f x$ encode colored HOMFLY-PT polynomials of the unknot in framing $f$ ($f=3$ in the example in the figure).}  \label{fig-framed-unknot}
\end{figure}

\begin{figure}[h]
  \centering
  \includegraphics[width=0.8\textwidth]{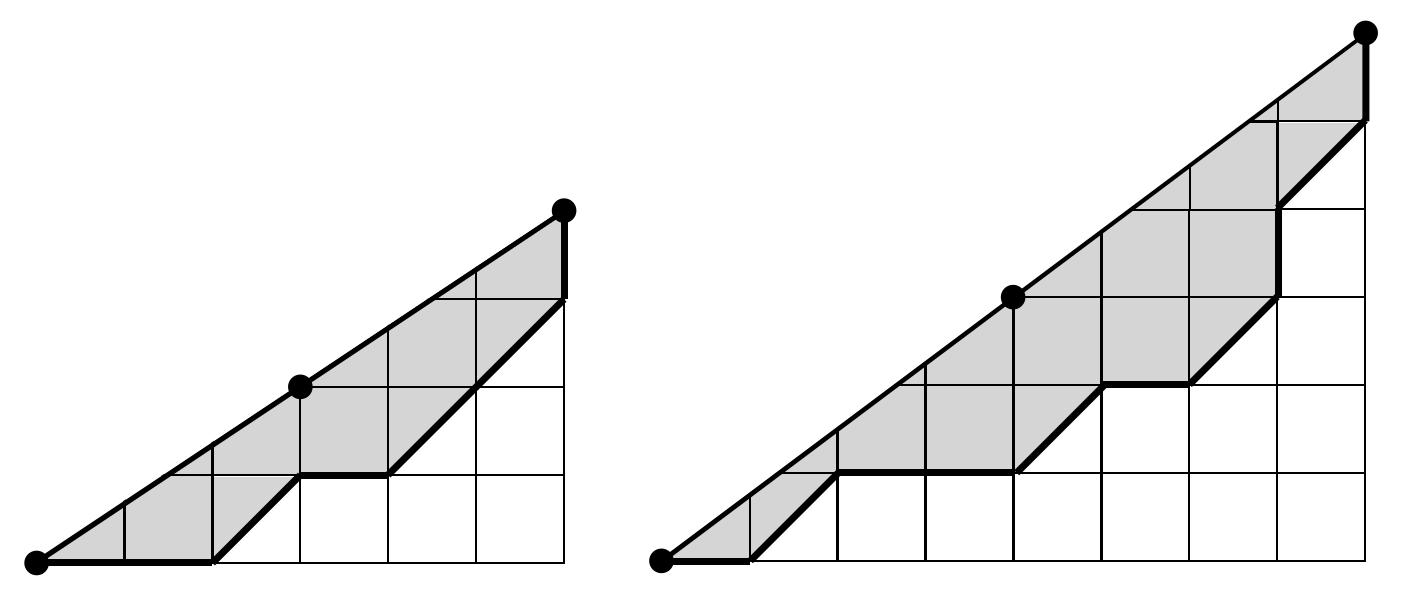}
  \caption{Generalized Schr{\"o}der paths under the line $y=\frac23 x$ (left) and $y=\frac34 x$ (right) encode respectively colored HOMFLY-PT polynomials of $(2,3)$ and $(3,4)$ torus knots.}  \label{fig-23-34}
\end{figure}

We call the lattice paths of our interest as generalized Schr{\"o}der paths. We show that they encode colored HOMFLY-PT polynomials $P_r(a,q)$ of a $(m,n)$ torus knot (also denoted $T_{m,n}$), in framing $mn$, colored by symmetric representations $S^r$.  Generalized Schr{\"o}der paths are paths in a square lattice composed of three possible steps: horizontal $(1,0)$,  vertical $(0,1)$, and diagonal ones $(1,1),$ which start in the origin. 
The type $(m,n)$ of a torus knot under consideration determines the region in which the paths lie: we consider paths in the positive quadrant of a plane and below the line $y=\frac{m}{n}x$ of slope $\frac{m}{n}$. The dependence on $a$ is captured by the number of diagonal steps in a given path, while the dependence on $q$ by the area between the path and the line $y=\frac{m}{n}x$. In literature, the paths under the line of unit slope, $y=x$, made of the above three basic steps, are referred to as Schr{\"o}der paths. By generalized Schr{\"o}der paths we mean paths that lie below a line of an arbitrary rational slope $\frac{m}{n}$, see examples in fig. \ref{fig-unknot},  \ref{fig-framed-unknot} and \ref{fig-23-34}.  
In particular, paths under the line of the slope $\frac1f$ capture invariants of the unknot in framing $f$.

Combinatorics of lattice paths is a broad and important research direction, within which various types of lattice paths are studied \cite{Bizley_paths,Duchon,banderier200237,Banderier-Wallner}. In particular, Schr{\"o}der paths are generalizations of Dyck paths. Dyck paths are made of only two elementary steps: horizontal $(1,0)$, and vertical $(0,1)$ (without the diagonal step).  The number of Dyck paths from the origin to the point $(k,k)$ in the square lattice is equal to the Catalan number $C_k$, see fig. \ref{fig-unknot} (right).  Another famous combinatorial objects are Duchon paths, which are paths consisting also of only horizontal and vertical steps, which lie below the line $y=\frac23 x$. By generalized Duchon paths we mean paths made of horizontal and vertical steps, which lie below the line of arbitrary rational slope $y=\frac{m}{n} x$ -- in other words, these are Schr{\"o}der paths without diagonal steps. All these types of paths also play a role in our analysis.

While various properties of lattice paths are known, their relations to knot invariants and related physical concepts are rather new and have not been deeply explored.  Some initial results in this context are presented in the previous work that we coauthored \cite{Panfil:2018sis},  which shows that generalized Duchon paths (i.e. those that do not involve diagonal steps) encode extremal colored HOMFLY-PT polynomials (i.e. suitably defined $a$-independent part of full colored HOMFLY-PT polynomials) of torus knots. We also found that Schr{\"o}der paths (those under the line $y=x$, of the unit slope) capture colored HOMFLY-PT invariants of the (unframed) unknot.  In the current work we generalize these results to full colored HOMFLY-PT polynomials of arbitrary torus knots and show that their $a$-dependence is simply captured by including diagonal steps in the construction of paths. 

Our work also reveals links between combinatorics of lattice paths and quiver representation theory.  A prominent role in this relation is played by the knots-quivers correspondence \cite{Kucharski:2017poe,Kucharski:2017ogk},  which enables to express colored HOMFLY-PT polynomials of torus knots in terms of quiver generating series with appropriate identification of generating parameters.  In this paper we show that another identification of parameters in the quiver generating series (for the same quiver) yields generating functions of corresponding generalized Schr{\"o}der paths.  This relation also implies that certain elementary paths are in one-to-one correspondence with generators of uncolored HOMFLY-PT homology, thereby generalizing our considerations to the realm of such homologies \cite{DGR, Gukov:2011ry,Gorsky:2013jxa}.  In fact, quivers that arise in this context correspond to unreduced invariants of $(m,n)$ torus knots, in framing $mn$. We explicitly identify and list these quivers for a few interesting examples. In particular, we determine the full quivers (both reduced and unreduced one,  encoding the full $a$-dependence) for $(3,4)$ torus knot (i.e. $8_{19}$ knot).  It follows that the whole information about the counts of a given class of generalized Schr{\"o}der paths is encoded in a finite set of parameters, i.e.  the entries of a quiver matrix $C$ and parameters that determine the specialization of quiver generating parameters $x_i$.

To sum up, for a quiver corresponding to a $(m,n)$ torus knot, two different identifications of parameters produce either colored HOMFLY-PT polynomials for this knot, or generating series of generalized Schr{\"o}der paths under the line of the slope $\frac{m}{n}$.  Therefore, the relation between knot polynomials and counts of Schr{\"o}der paths is not completely straightforward (as it involves these two parameter specializations).  However, we also find a direct expressions for the counts of Schr{\"o}der paths in terms of knot polynomials. Amusingly, it involves superpolynomials of colored quadruple knot homology, introduced in \cite{Gorsky:2013jxa} and denoted $\P^Q_r(a,q,t_r,t_c)$: we find a specialization of parameters of such polynomials (different than the one that produces colored HOMFLY-PT polynomials) that immediately yields counts of Schr{\"o}der paths.

Finally,  it is known that generating functions of lattice paths (in $q=1$ limit) should satisfy algebraic equations.  This statement also follows from their relation to colored knot polynomials,  which take form of $q$-holonomic functions and for this reason satisfy recursion relations, reducing in $q=1$ limit to algebraic equations. We explicitly determine such algebraic equations, as well as related recursion relations that capture the dependence on $q$. In the knot theory context such relations are called classical and quantum (generalized) A-polynomials \cite{AVqdef,FGS,superA,FGSS,Fuji:2013rra,Nawata}, hence we also refer to such objects as A-polynomials (for lattice paths).  We provide a number of nontrivial examples of such A-polynomials.

We believe our results deserve further analysis and generalizations.  They provide new direct links between knot theory, quiver representation theory and combinatorics, so it would be important to prove them using and relating to each other the methods of these fields. Furthermore, while in general the counting interpretation of knot invariants follows from considerations in physics, interpretation of certain objects as BPS states and engineering of brane systems in string theory, it would be instructive to rederive our results from physical setup more explicitly.  The relations to lattice paths that we present could be also generalized in various ways, e.g. to other values of framing, to knots (and links) other than torus knots, to the refined case, to the categorified level, etc.  Such relations might involve other combinatorial models than just those of lattice paths. It would be also interesting to relate our results to the relation between quantum A-polynomials and combinatorics on words presented in \cite{Kucharski:2016rlb}.

The plan of the paper is as follows.  In section \ref{sec-review} we summarize various features of knot homologies and the knots-quivers correspondence. In section \ref{sec-lattice} we present in general terms the main results of this paper, i.e. the relations between generating functions of lattice paths, colored polynomials for torus knots, and quiver generating series. In section  
\ref{sec-quivers} we illustrate statements from the previous section in explicit examples related to a few particular torus knots and generalized Schr{\"o}der paths. In section \ref{sec-A-results} we determine explicit form of A-polynomials for several classes of generalized Schr{\"o}der paths. 


\section{Knot homologies and knots-quivers correspondence}    \label{sec-review}
 
In this section we recall basic facts and summarize the notation concerning colored knot polynomials associated to various knot homologies, as well as the knots-quivers correspondence.
 
\subsection{Colored superpolynomials of quadruply-graded homology}    \label{ssec-PQr}
 
In this work we relate generating functions of lattice paths to colored HOMFLY-PT polynomials for torus knots $P_r(a,q)$, where $r$ denotes the coloring by the symmetric representation $S^r$. The  polynomials $P_r(a,q)=\P_r(a,q,-1)$ are obtained as $t=-1$ specialization of colored superpolynomials of HOMFLY-PT homology $\P_r(a,q,t)$ \cite{DGR, Gukov:2011ry}. However, it is of advantage to consider yet more general knot polynomials, namely colored superpolynomials of quadruply-graded homologies introduced in \cite{Gorsky:2013jxa}, denoted $\P^Q_r(a,q,t_r,t_c)$. In the convention that we use here, the specialization of $\P^Q_r(a,q,t_r,t_c)$ with $t_r=1$ and $t_c=t$ yields the above colored superpolynomials\footnote{Another consistent choice often considered in literature, e.g. in \cite{Kucharski:2017ogk, Gukov:2011ry}, is to define superpolynomials as $t_r=t,t_c=1$ specialization of quadruply-graded homology, i.e. $P_r(a,q,t) = \P^Q_r(a,q,t,1)$.} 
\begin{equation}
\P_r(a,q,t) = \P^Q_r(a,q,1,t).   \label{Pr-PrQ}
\end{equation}
On the other hand, as we will see in what follows, generalized Schr\"oder paths can be also related directly to the specialization 
$\P^Q_r(a,q=1,t_r=1,t_c=q^{-1})\equiv \P^Q_r(a,1,1,q^{-1})$. To start with, we summarize explict expressions for colored superpolynomials of quadruply-graded homology for several knots, which we will take advantage of in what follows.  Invoking the expressions for $q$-Pochhamer and $q$-binomial series
\begin{equation}
(x;q)_k = \prod_{j=0}^{k-1} (1-xq^j),\qquad {r \brack k} _q= \frac{(q;q)_r}{(q;q)_k (q;q)_{r-k}},
\end{equation}
we have

\medskip
$\bullet$ Trefoil knot (i.e. $(2,3)$ torus knot, $T_{2,3}$, or $3_1$ knot):
\begin{equation}
\P^Q_r(a,q,t_r,t_c)(3_1)=a^{2r} q^{-2r}  \sum_{k=0}^r {r \brack k}_{q^{2} t_c^{2}} q^{2(r+1)k} t_r^{2k} t_c^{2r k} \prod_{i=1}^{k} (1+ a^2 q^{2(i-2)} t_r t_c^{2i-1}), \label{PQr-trefoil}
\end{equation}
which can be also rewritten (for simplicity, including only the dependence on $t_c$) as
\begin{align}
\begin{split}
\P^Q_r(a,q,1,t_c)(3_1) &= \sum_{k=0}^r { r \brack k}_{q^2 t_c^2}  a^{2k} q^{2(k^2-k)} t_c^{2k^2} \times \\
& \quad \times (-1)^{r-k} a^{2k-2r} q^{2r(k-r)+(r-k)(r-k+1)} t_c^{2r(k-r)+(r-k)(r-k-1)} \times \\
& \quad \times \prod_{i=1}^k (1+ a^{-2} q^{2(2-i)}  t_c^{1-2i})(1+ a^{-2} q^{2(1-r-i)}  t_c^{1-2i-2r}).
\end{split}
\end{align}

$\bullet$ Torus knots $(2,2p+1)$ (i.e.  $T_{2,2p+1}$ or $(2p+1)_1$ knots) for arbitrary $p$:
\begin{align}
\begin{split}
& \P^Q_r(a,q,t_r,t_c)(T_{2,2p+1}) = a^{2pr} q^{-2pr}   \sum_{0\le k_p\le k_{p-1}\le \cdots\le k_2\le k_1\le r} {r \brack k_1}_{q^{2} t_c^{2}} {k_1 \brack k_2}_{q^{2} t_c^{2}} \cdots {k_{p-1} \brack k_p}_{q^{2} t_c^{2}} \\
& \ \times q^{2(2r+1)\sum_{i=1}^p k_i -2\sum_{i=1}^p k_{i-1}k_i} t_r^{2\sum_{i=1}^p k_i} t_c^{4r \sum_{i=1}^p k_i -2\sum_{i=1}^p k_{i-1}k_i} \prod_{i=1}^{k_1} (1+ a^2 q^{2(i-2)} t_r t_c^{2i-1})  \label{Pr-22p1}
\end{split}
\end{align}
with the convention $k_0=r$. The expression for trefoil arises for $p=1$, for $(2,5)$ torus knot (i.e. $5_1$ knot) for $p=2$, etc.

\medskip

$\bullet$ Figure-eight knot, $4_1$:
\begin{align}
\begin{split}
\P^Q_r(a,q,t_r,t_c)(4_1) & =\sum_{k=0}^r { r \brack k}_{q^2 t_c^2}  a^{2k} q^{2(k^2-k)} t_r^{2k} t_c^{2k^2}\times \\
& \times  \prod_{i=1}^k (1+ a^{-2} q^{2(2-i)} t_r^{-1} t_c^{1-2i})(1+ a^{-2} q^{2(1-r-i)} t_r^{-3} t_c^{1-2i-2r}).   \label{P-41}
\end{split}
\end{align}

$\bullet$ Knot ${5_2}$:
\begin{align}
\begin{split}
\P^Q_r(a,q,t_r,t_c)(5_2)&=a^{2r} q^{-2r}  \sum_{k=0}^r \sum_{j=0}^k {r \brack k}_{q^{2} t_c^{2}} {k \brack j}_{q^{2} t_c^{2}} \times \\
& \quad \times a^{2j+2k} t_r^{3k+2j}q^{k^2-k+2j^2-2j+2rk}  t_c^{k^2+2j^2+2r k}\times \\
&\quad \times(-a^{-2}q^2 t_r^{-1}t_c^{-1};q^{-2}t_c^{-2})_k (-a^{-2}q^{-2r} t_r^{-3}t_c^{-2r-1};q^{-2}t_c^{-2})_j.   \label{P-52}
\end{split}
\end{align}

$\bullet$ $(3,4)$ torus knot (i.e. $8_{19}$ knot)  \cite{Gukov:2015gmm}:
\begin{align}
\begin{split}
\P^Q_r(a,q,t_r,t_c)(T_{3,4})&=a^{6 r}q^{4r^2-4r}t_r^{4r} t_c^{4 r^2}\sum _{j=0}^r  \sum _{j\ge k_1\ge k_2\ge k_3\ge 0} {r\brack j}_{q^2 t_c^2}{j\brack k_1}_{q^2 t_c^2}{k_1\brack k_2}_{q^2 t_c^2}{k_2\brack k_3}_{q^2 t_c^2}  \\
&  \times q^{-2 j + 2k_3-2k_2k_3+2(k_1+k_2+k_3)r+j(k_2+k_3-2r)}\times\\
& \times t_r^{2 (-2 j + k_1 + k_2 + k_3 )} t_c^{2 ( -k_1 -k_2 -k_2 k_3  +(k_1+ k_2  + k_3) r + j (k_2 + k_3 - 2 r) )} \times\\
& \times  \left(-a^2 q^{-2}t_r t_c;q^2 t_c^2\right)_{r-j} \left(-a^2 q^{2r} t_r^3 t_c^{1+2 r} ;q^2 t_c^2\right)_{r-j}  \times\\
&\times \left(-a^2q^{2(r-j-1)}t_r t_c^{1+2 (r-j)} ;q^2 t_c^2\right)_{k_1} .    \label{Psuper819}
\end{split}
\end{align}



\subsection{Knots-quivers correspondence}

A crucial role in this work is also played by the knots-quivers correspondence \cite{Kucharski:2017poe,Kucharski:2017ogk,Stosic:2024egq}. This correspondence relates symmetrically colored HOMFLY-PT polynomials of a knot $K$ to quiver generating series for the corresponding (symmetric) quiver $Q_K$. Explicitly, the correspondence states that all the information about all symmetrically-colored HOMFLY-PT polynomials $P_r(K)\equiv P_{r}(a,q)(K)$ of the knot $K$ is contained in the integral symmetric matrix $C$ of size $k\times k$, corresponding to the adjacency matrix of the quiver $Q_K$ with $k$ vertices, as well as two integral vectors $\mathbf{a}=(a_1,\ldots,a_k)$ and $\mathbf{q}=(q_1,\ldots,q_k)$, which encode homological $(a,q)$-degrees of $k$ generators of uncolored HOMFLY-PT homology. For the reduced HOMFLY-PT polynomials the relationship reads
\begin{equation}
\label{kqc}\sum_{r\ge 0}\frac{P_{r}(a,q)(K)}{(q^2;q^2)_r}x^r=
P_C(x_1,\ldots,x_k)\mid_{x_i=a^{a_i}q^{q_i-C_{i,i}}x},  \end{equation}
where the quiver generating series $P_C(x_1,\ldots,x_k)$ has the form
\begin{equation}\label{kqc1}
P_C(x_1,\ldots,x_k)=\sum_{d_1,\ldots,d_{k}}(-q)^{\sum_{i,j}C_{i,j}d_id_j}
\frac{x_1^{d_1} x_2^{d_2}\ldots x_k^{d_k}}{\prod_{i=1}^{k}(q^2;q^2)_{d_i}}.
\end{equation}

A quiver matrix for the mirror image to a given knot is given by $I_{k\times k}-C$, where $C$ is a quiver matrix for the original knot. Furthermore, a quiver matrix corresponding to colored HOMFLY-PT polynomials with a framing shifted by $f\in\Z$ (with respect to polynomials corresponding to the matrix $C$) takes form
\begin{equation}
C+f\left[\begin{array}{ccc}1&\cdots&1\\
\vdots&\ddots&\vdots\\
1&\cdots&1\end{array}\right].
\end{equation}



In this work we are primarily interested in generating series of unreduced knot polynomials, as they turn out to be directly related to the counting of lattice paths. Unreduced colored HOMFLY-PT polynomials are defined by
\begin{equation}
\bar{P}_r(K) \equiv \bar{P}_r(a,q)(K)=P_r(K)\bar{P}_r(\textrm{unknot})= P_r(K) a^{-r}q^r\frac{(a^2;q^2)_r}{(q^2;q^2)_r},  \label{Pr-unreduced}
\end{equation} 
and their generating series can also be written in the form of quiver generating series for an appropriate quiver matrix $\bar{C}$
\begin{equation}
\label{kqc2}
\sum_{r\ge 0}\bar{P}_{r}(K)x^r=
P_{\bar{C}}(x_1,\ldots,x_k)\mid_{x_i=a^{a_i}q^{q_i-\bar{C}_{i,i}}x},
\end{equation}
where again
\begin{equation}
\label{kqc1unreduced}
P_{\bar{C}}(x_1,\ldots,x_k)=\sum_{d_1,\ldots,d_{k}}(-
q)^{\sum_{i,j}\bar{C}_{i,j}d_id_j}
\frac{x_1^{d_1} x_2^{d_2}\ldots x_k^{d_k}}{\prod_{i=1}^{k}(q^2;q^2)_{d_i}}.
\end{equation}
As shown in \cite{Kucharski:2017ogk}, $\bar{C}$ is simply related to $C$, as we recall now. First, we rewrite (\ref{kqc}) as follows
\begin{equation}
\label{form1}
\sum_{r\ge 0}P_{r}(K)x^r=\sum_{d_1,\ldots,d_{k}}(-
q)^{\sum_{i,j}C_{i,j}d_id_j}a^{\sum_ia_id_i}q^{\sum_i(q_i-C_{ii})d_i}
\frac{(q^2;q^2)_{\sum_i d_i}}{\prod_{i=1}^{k}(q^2;q^2)_{d_i}}x^{\sum_i d_i}.
\end{equation}
In terms of unreduced polynomials (\ref{Pr-unreduced}) and using the relation $r=\sum_i d_i$ we get
\begin{align}
\begin{split}
&\sum_{r\ge 0}\bar{P}_{r}(K)x^r= \\
&= \sum_{d_1,\ldots,d_{k}}(-q)^{\sum_{i,j}C_{i,j}d_id_j}a^{\sum_ia_id_i}q^{\sum_i(q_i-C_{ii})d_i}a^{-{\sum_i d_i}}q^{\sum_i d_i}\frac{(a^2;q^2)_{\sum_i d_i}}{\prod_{i=1}^{k}(q^2;q^2)_{d_i}}x^{\sum_i d_i}.
\end{split}
\end{align}
Using now
\begin{equation}
(a^2;q^2)_r=(a^2;q^2)_{d_1+\cdots+d_k}=(a^2;q^2)_{d_1}(a^2q^{2d_1};q^2)_{d_2}\cdots(a^2q^{2(d_1+\cdots+d_{k-1})};q^2)_{d_k}
\end{equation}
and the $q$-binomial identity 
\begin{equation}
\frac{(y^2;q^2)_d}{(q^2;q^2)_d}=\sum_{\alpha
+\beta=d}(-1)^{\alpha}y^{2\alpha}q^{\alpha^2-\alpha}\frac{1}{(q^2;q^2)_{\alpha}
(q^2;q^2)_{\beta}}
\end{equation}
we get 
\begin{align}
\begin{split}
& \sum_{r\ge 0}\bar{P}_{r}(K)x^r=\!\!\!\!\!\sum_{\begin{array}{c}\alpha_1,\ldots,\alpha_{k}\\
\beta_1,\ldots,\beta_k\end{array}}(-
q)^{\sum_{i,j}C_{i,j}(\alpha_i+\beta_i)(\alpha_j+\beta_j)}a^{\sum_i(a_i-1)
(\alpha_i+\beta_i)}q^{\sum_i(q_i-C_{ii}+1)(\alpha_i+\beta_i)}\times \\
&\quad  \times(-1)^{\sum_i\alpha_i}  a^{2\sum_i\alpha_i}q^{\sum_i(\alpha_i^2-\alpha_i)}q^{2\sum_i\alpha_i(\alpha_1+
\beta_1+\cdots+\alpha_{i-1}+\beta_{i-1})}
\frac{x^{\sum_i{\alpha_i}+\sum_i\beta_i}}{\prod_{i}(q^2;q^2)_{\alpha_i}
(q^2;q^2)_{\beta_i}}.
\end{split}
\end{align}
Finally, concatenating variables $\alpha$ and $\beta$ 
\begin{equation}
(\gamma_1,\ldots,\gamma_{2k}):=(\alpha_1,\beta_1,\alpha_2,\beta_2,\ldots,
\alpha_k,\beta_k),
\end{equation}
we get 
\begin{equation}\label{form2}\sum_{r\ge 0}\bar{P}_{r}(K)x^r=\sum_{\gamma_1,\ldots,\gamma_{2k}}(-
q)^{\sum_{i,j}\bar{C}_{i,j}\gamma_i\gamma_j}a^{\sum_i\bar{a}_i\gamma_i}q^{\sum_i(\bar{q}_i-\bar{C}_{ii})\gamma_i}
\frac{x^{\sum_i\gamma_i}}{\prod_{i=1}^{2k}(q^2;q^2)_{\gamma_i}},\end{equation}
where 
$\bar{C}$ is $2k\times 2k$ matrix
\arraycolsep 4pt
\begin{equation}
\bar{C}=C\otimes\left[\begin{array}{cc}1&1\\1&1\end{array}\right]+
\left[\begin{array}{ccccccccc}
1&0&1&0&1&0&\cdots&1&0\\
0&0&1&0&1&0&\cdots&1&0\\
1&1&1&0&1&0&\cdots&1&0\\
0&0&0&0&1&0&\cdots&1&0\\
1&1&1&1&1&0&\cdots&1&0\\
0&0&0&0&0&0&\cdots&1&0\\
\vdots&\vdots&\vdots&\vdots&\vdots&\vdots&\ddots&\vdots&\vdots\\
1&1&1&1&1&1&\cdots&1&0\\
0&0&0&0&0&0&\cdots&0&0\end{array}\right]
\end{equation}
and 
\begin{align}
\begin{split}
\mathbf{\bar{a}}&=(a_1+1,a_1-1,a_2+1,a_2-1,\ldots,a_k+1,a_k-1), \\
\mathbf{\bar{q}}&=(q_1+1,q_1+1,q_2+1,q_2+1,\ldots,q_k+1,q_k+1).
\end{split}
\end{align}


\section{Lattice paths from knots, quivers and A-polynomials}    \label{sec-lattice}

In this section we present the main results of this work, namely explicit relations between generating functions of lattice paths and colored polynomials for torus knots, as well as quiver generating series. Furthermore, we also show that generating functions of lattice paths satisfy algebraic equations that we call A-polynomials, as well as recursion relations captured by operators called quantum A-polynomials (in analogy with A-polynomials for knots \cite{AVqdef,FGS,superA,FGSS,Fuji:2013rra,Nawata}).  Explicit examples and illustrations of these statements will be presented in the following sections.

As already explained in the introduction, lattice paths of our interest are generalized Schr\"oder paths, which consist of elementary steps to the right, upwards, and a diagonal one, in a square lattice. We count paths that start at the origin, are located above the horizontal axis, and below a line of rational slope, $y=\frac{m}{n}x$. We show that the count of such paths can be obtained from the generating series of the colored HOMFLY-PT polynomials of $mn$-framed $(m,n)$-torus knot, so that the powers of variable  $a$ in the colored HOMFLY-PT polynomials count the number of diagonal steps in a Schr\"oder path, while powers of $q$ capture the area between a given path and the line $y=\frac{m}{n}x$. This framework naturally generalizes the results from \cite{Panfil:2018sis} where it was shown that the lattice paths without diagonal steps correspond to the generating series of the bottom rows of the colored HOMFLY-PT polynomials of torus knots, which is essentially $a=0$ specialization of our current results.


 \subsection{Generalized Schr\"oder paths from quivers}
 
Our first result relates generalized Schr\"oder paths under the line of rational slope $\frac{m}{n}$ to quiver generating series,  for a quiver corresponding to $(m,n)$ torus knot. 
 
 Let $m$ and $n$ be mutually prime positive integers, such that $m< n$. Let $K$ be a left handed $
 (m,n)$ torus knot with framing $f=mn$. Let $\bar{C}$ be the matrix of the quiver $Q$, 
 corresponding to the unreduced symmetrically colored HOMFLY-PT polynomials of the 
 knot $K$.  Let $\mathbf{a}=(a_1,a_2,\ldots,a_{2k})$ be a vector corresponding 
 to the $a$-degrees of the nodes of the quiver $Q$. Let $a_{\max}:=\max_i\{a_i\}$, $i=1,\ldots,2k$, and  $a'_i=a_{\max}-a_i$, $i=1,\ldots,2k$.  Furthermore, let $P_{\bar{C}}(x_1,\ldots,x_{2k})$ be the generating series for the quiver $Q$
\begin{equation}
P_{\bar{C}}(x_1,\ldots,x_{2k})=\sum_{d_1,\ldots,d_{2k}}(-q)^{\sum_{i,j}\bar{C}_{i,j}d_id_j}\frac{\prod_{i=1}^{2k}x_{i}^{d_i}}{\prod_{i=1}^{2k}(q^2;q^2)_{d_i}}
\end{equation}
and let $\bar{P}_{\bar{C}}(x)$ be the following one variable series specialization of $P_{\bar{C}}(x_1,\ldots,x_{2k})$
\begin{equation}
x_i \mapsto (-1)^{f}(-a)^{a'_i}x,\quad i=1,\ldots,2k.   \label{xi-x}
\end{equation}
Finally,  let 
\begin{equation}
y(x,a,q) = \frac{\bar{P}_{\bar{C}}(qx)}{\bar{P}_{\bar{C}}(q^{-1}x)}=\sum_{l\ge 0}N_l(a,q)x^l=\sum_{i,j,l}n_{i,j,l}a^iq^jx^l.     \label{prop1}
\end{equation}
 
\begin{proposition}
The number $n_{i,j,l}$ introduced above equals the number of generalized Schr\"oder paths  under the line $y=\frac{m}{n}
 x$ that start at $(0,0)$ and end at $(nl,ml)$, with $\frac{i}{2}$ diagonal steps and such that the area between the path and the line  $y=\frac{m}{n} x$ is equal to $\frac{j}{2}$. In particular, for each $l\ge 0$, only finitely many coefficients $n_{i,j,l}$ are non-zero, i.e. for each $l\ge 0$, $N_l(a,q)\in\N[a^{\pm 1},q^{\pm 1}]$ is a (Laurent) polynomial.    \label{prop1txt}
\end{proposition}

In other words, each diagonal step in a given path contributes a factor of $a^2$ to its weight,  and the area of each unit lattice square that contributes to the weight of a path (i.e. a square between a given path and the line $y=\frac{m}{n}x$) is $q^2$ -- this is a consequence of our convention that involves $a^2$ and $q^2$ in expressions for knot polynomials and quiver generating series. In the $q=1$ limit the above result reduces to the generating function of generalized Schr\"oder paths without weighting them by the area
\begin{equation}
y(x,a)=\lim_{q\to 1} \frac{\bar{P}_{\bar{C}}(q x)}{\bar{P}_{\bar{C}}(q^{-1}x)}= \sum_{l\ge 0}N_l(a,1)x^l .  \label{yxa}
\end{equation}


\subsection{Generalized Schr\"oder paths from quadruply-graded knot homology}    \label{ssec-quadruply}

The above proposition \ref{prop1txt} expresses generating series of lattice paths in terms of quiver generating series, for a quiver associated to a torus knot. However, the identification of variables (\ref{xi-x}) is different than the one in (\ref{kqc}) that yields colored HOMFLY-PT polynomials. Nonetheless, we find that expressions for generalized Schr\"oder numbers can be obtained directly from colored invariants of torus knots -- however, to achieve this we need to consider expressions for superpolynomials of quadruply-graded homologies.

To this end, let $\P_r(a,q,t)$ be the superpolynomial of the reduced colored  HOMFLY-PT homology of the knot $K$. It can be obtained as a specialization (\ref{Pr-PrQ}) of superpolynomial of quadruply-graded homology,  $\P_r(a,q,t)=\P^Q_r(a,q,1,t)$. Recall we provided explicit expressions for $\P^Q_r(a,q,t_r,t_c)$ for some knots in section \ref{ssec-PQr}. For $(m,n)$ torus knot we also introduce another specialization and the following notation  
\begin{align}
\begin{split}
p'_r(a,t)&= a^{-(m-1)(n-1)r} \P_r(a,q=1,t),\\
\bar{p}_r(a,q)&= p'_r(a,t=q^{-1})=a^{-(m-1)(n-1)r} \P_r(a,q=1,t=q^{-1}),\label{for11}\\
p_r(a,q)&=q^{ fr^2} \bar{p}_r(a,q)=a^{-(m-1)(n-1)r} q^{f r^2} \P_r(a,q=1,t=q^{-1}),
\end{split}
\end{align}
where a factor $a^{-(m-1)(n-1)r}$ simply shifts the $a$-degrees, so that the terms with the lowest $a$-degree in $\P_r(a,q,t)$ have $a$-degree equal to zero.  We then consider the following specialization of the generating series of unreduced superpolynomials shifted by the above factor $a^{-(m-1)(n-1)r}$
\begin{align}
\begin{split}
\bar{P}(x,a,q)&= 1+\sum_{r\ge 1}(-1)^r\frac{\prod_{i=1}^r (a^2+q^{2i-1})}{(q^2;q^2)_r} q^{f r^2}a^{-(m-1)(n-1)r}\P_r(a,q=1,t=q^{-1}) x^r =  \\ 
&=1+\sum_{r\ge 1}(-1)^r\frac{\prod_{i=1}^r (a^2+q^{2i-1})}{(q^2;q^2)_r} p_r(a,q) x^r,\label{Pxaq}
\end{split}
\end{align}
and to make contact with the counting of lattice paths corresponding to $(m,n)$ torus knot we also fix the framing as $f=mn$.

Finally, we take the quotient that governs the growth of $p_r(a,q)$ 
\begin{equation}
y(x,a,q)=\frac{\bar{P}(q x,a,q)}{\bar{P}(q^{-1}x,a,q)} = 
\sum_{l\ge 0}N_l(a,q)x^l=\sum_{i,j,l}n_{i,j,l}a^iq^jx^l . \label{yxaq}
\end{equation}

\begin{proposition}
This $y(x,a,q)$ agrees with (\ref{prop1}); in particular the number $n_{i,j,l}$ in (\ref{yxaq}) is the same number that appears in (\ref{prop1}), i.e. it equals the number of generalized Schr\"oder paths  under the line $y=\frac{m}{n} x$ that start at $(0,0)$ and end at $(nl,ml)$, with $\frac{i}{2}$ diagonal steps and such that the area between the path and the line  $y=\frac{m}{n} x$ is equal to $\frac{j}{2}$.     \label{prop2txt}
\end{proposition}



\subsection{Basic Schr{\"o}der paths and homology generators}    \label{sec-A}

Let us call Schr{\"o}der paths under the line of the slope $\frac{m}{n}$ from the origin to the point $(n,m)$ as the basic paths. One interesting consequence of the above statements is the following:

\begin{proposition}
The number of basic paths weighted by the number of diagonal steps agrees with the number of unreduced generators of uncolored HOMFLY-PT homology of the $(m,n)$ torus knot, weighted by the $a$-degree.     \label{paths-generators}
\end{proposition}
Indeed,  it follows from (\ref{yxaq}) that to first order in $x$
\begin{equation}
y(x,a,q) = 1 + \P_1(a,q=1,t=q^{-1}) q^{f-1} (a^2 + q) x + \mathcal{O}(x^2)
\end{equation}
and thus
\begin{equation}
y(x,a,q=1) = 1 + \P_1(a,q=1,t=1)  (a^2 + 1) x + \mathcal{O}(x^2).
\end{equation}
The term of order $x$ captures counting of basing paths with each diagonal step weighted by $a^2$.  At the same time,  this term  represents the number of reduced uncolored HOMFLY-PT generators weighted by $a$ (which are captured by $\P_1(a,q=1,t=1)$) and analogous term weighted by extra $a^2$; these two terms indeed represent all unreduced HOMFLY-PT generators.

In consequence, one could interpret Schr{\"o}der paths of arbitrary length as being constructed from basic paths, in analogy to the construction of colored HOMFLY-PT homology from the uncolored one \cite{Gukov:2011ry}.


\subsection{A-polynomials for lattice paths}   

A-polynomials are algebraic curves well known in knot theory. The original A-polynomial was defined through $SL(2,\mathbb{C})$ character variety of knot complement \cite{CCGLS}. Later it was realized that it is related to the asymptotic behaviour of colored Jones polynomial via famous AJ-conjecture \cite{gelca, garoufalidis, Gukov:2003na}. 
More precisely, in this latter sense, A-polynomials arise in two versions, classical and quantum.
Classical A-polynomials encode asymptotics of colored knot polynomials for large color, while quantum A-polynomials provide recursion relations for colored knot polynomials (for all colors, not only large color). Such A-polynomials were also generalized to colored HOMFLY-PT polynomials, related to augmentation polynomials \cite{FGS}, and further generalized to the refined case \cite{FGS} and to super-A-polynomials \cite{superA,FGSS,Fuji:2013rra}. 

We introduce now analogous objects, which we call quantum and classical A-polynomials for lattice paths and which capture information about generating functions of such paths that follow from expressions of the form (\ref{Pxaq}). First, using similar techniques as for knots \cite{superA,FGSS,Fuji:2013rra}, we determine quantum A-polynomials $\widehat{A}(\hat x, \hat y)$, which are operators that encode difference equations satisfied by (\ref{Pxaq}).  Such operators can be determined as follows.  Let us write the expression of the form (\ref{Pxaq}) as $\bar{P}(x,a,q)\equiv \bar{P}(x) =\sum_{r=0}\bar{P}_r x^r$.  The coefficients $\bar{P}_r$ are $q$-holonomic (i.e. they are given as sums of expressions that involve at most quadratic powers of $q$, $q$-Pochhammers, and  $q$-binomials), so from a general theory \cite{doi:10.1080/10236190701264925,Garoufalidis:2016zhf} it follows that they satisfy recursion relation of the form
\begin{equation}
\alpha_0(q^r,q) \bar{P}_r + \alpha_1(q^r,q) \bar{P}_{r+1} + \ldots \alpha_k (q^r,q) \bar{P}_{r+k}  = 0.     \label{A-recursion}
\end{equation}
In practice,  such recursion relations can be found using e.g. \emph{qZeil} package \cite{qZeil}.  The relation (\ref{A-recursion}) can be equivalently written as
\begin{equation}
\mathsf{\widehat{A}(\hat x,\hat y)} \bar{P}_r  = 0,\quad \textrm{for} \quad   \mathsf{\widehat{A}(\hat x,\hat y)}  =  \sum_{j=0}^k  \alpha_j(\mathsf{\hat{x}},q) \mathsf{\hat{y}}^j     \label{A-recursion-2}
\end{equation}
where $\mathsf{\hat{x}}$ and $\mathsf{\hat{y}}$ act respectively as multiplication by $q^r$ and $\mathsf{\hat{y}}\bar{P}_r=\bar{P}_{r+1}$.  This recursion relation can be easily turned into a relation for the generating function $\bar{P}(x)=\sum_{r}\bar{P}_r x^r$  \cite{Garoufalidis:2015ewa}.  Indeed acting with $\mathsf{\hat{y}}$ or $\mathsf{\hat{x}}$ on the whole generating series 
\begin{equation}
\mathsf{\hat{y}} \sum_r \bar{P}_r x^r = \sum_r \bar{P}_{r+1} x^r = x^{-1} \sum_r \bar{P}_{r} x^r, \quad \mathsf{\hat{x}} \bar{P}(x) = \sum_r q^r \bar{P}_r  x ^r = \bar{P}(qx),
\end{equation}
has the same effect as respectively acting with $\hat{x}^{-1}$ or $\hat{y}^{1/2}$, whose action is defined as
\begin{equation}
\hat x \bar{P}(x) = x \bar{P}(x),\quad \hat y \bar{P}(x) = \bar{P}(q^2 x),\quad \hat y\hat x = q^2 \hat x\hat y. \label{hatx-haty}
\end{equation}
(The power of $1/2$ in the redefinition of $\mathsf{\hat x}$ into $\hat{y}^{1/2}$ is a consequence of our convention, in which $q$-Pochhammers have the argument $q^2$.) Therefore the quantum A-polynomial, defined as
\begin{equation}
\widehat{A}(\hat x,\hat y)  = \mathsf{\widehat{A}}(\hat{y}^{1/2},\hat{x}^{-1}),
\end{equation}
annihilates the generating series 
\begin{equation}
\widehat{A}(\hat x,\hat y) \bar{P}(x) = 0.
\end{equation}
Note that, when turning (\ref{A-recursion}) or (\ref{A-recursion-2}) into $\widehat{A}(\hat x,\hat y)$,  the term $\alpha_i(q^r,q)\bar{P}_{r+i}=\alpha_i(\mathsf{\hat{x}} ,q)\mathsf{\hat{y}}^i \bar{P}_{r}$ is turned into the term $\alpha_i(\hat{y}^{1/2},q) \hat{x}^{-i}\bar{P}(x)$. It is therefore useful to commute each factor $\hat{x}^{-i}$ to the left over $\alpha_i(\hat{y}^{1/2},q)$; this introduces additional $q^i$ each time $\hat{x}^{-i}$ passes over $\hat{y}^{1/2}$, and ultimately we can write $\widehat{A}(\hat x,\hat y)$ in the form 
\begin{equation}
\widehat{A}(\hat x,\hat y) = \sum_j \tilde{\alpha}_j(\hat{x},q) \hat{y}^j \equiv \sum_j \hat{x}^j \beta_j(\hat{y},q)  \label{qAhat}
\end{equation}
for appropriate coefficients $\tilde{\alpha}_j(\hat{x},q)$ and $\beta_j(\hat{y},q)$.  In the $q=1$ limit this quantum A-polynomial reduces to the classical A-polynomial
\begin{equation}
A(x,y) = 0, \quad  \textrm{for}\quad  A(x,y) =  \lim_{q\to 1} \widehat{A}(\hat x,\hat y),      \label{Axy}
\end{equation}
and this classical A-polynomial equation is satisfied by $y=y(x)$ defined as $q=1$ limit of (\ref{yxaq}) (or equivalently (\ref{yxa})). Such $y(x)$ captures asymptotics of (\ref{Pxaq}) for large $r$, and equivalently can be found by the saddle point method, which we now briefly review. 

To find $A(x,y)$ by the saddle point method, we determine first an equation $\mathsf{A(x,y)}$ that encodes asymptotic expansion of $\bar{P}_r$ for large $r$.  To this end we introduce $\mathsf{x}=q^r$ and $z_i=q^{k_i}$, where $k_i$ are summation variables that appear in (\ref{Pxaq}). Then we express $\bar{P}_r$ as
\begin{equation}
\bar{P}_r \sim \int \prod_{z_i} dz_i \exp \frac{1}{\hbar}\big( V(\mathsf{x},z_i)  + \mathcal{O}(\hbar)  \big).
\end{equation}
The saddle point analysis yields a system of equations
\begin{align}
\begin{split}
\mathsf{y} &= e^{\mathsf{x} \frac{\partial V(\mathsf{x} ,z_i)}{\partial \mathsf{x} } }, \\
1 &= e^{z_i \frac{\partial V(\mathsf{x} ,z_i)}{\partial z_i} }  , \quad \textrm{for each } i.    \label{saddle-eqs}
\end{split}
\end{align}
Eliminating all $z_i$ from the above set of equations we get a single relation $\mathsf{A(x,y)}=0$.  Then the A-polynomial equation we are interested in is obtained by the same change of variables as above: $\mathsf{x}\mapsto y^{1/2}, \mathsf{y}\mapsto x^{-1}$. In addition, we often rescale such a result by a simple overall factor (e.g. some power of $x$)
\begin{equation}
A(x,y) \sim \mathsf{A}(y^{1/2},x^{-1}), \label{Adual}
\end{equation}
so that we obtain a polynomial of the form $A(x,y)=1-y+\ldots$. This yields the result consistent with (\ref{Axy}) that was obtained as $q=1$ limit of quantum A-polynomial, i.e.  either (\ref{Axy}) and (\ref{Adual}) completely agree, or (\ref{Adual}) is one factor in (\ref{Axy}) (whose other factors are typically quite simple and are not relevant for our considerations of the classical limit).

Note that $V(\mathsf{x})\equiv V(\mathsf{x},z_i)$ that follows from (\ref{Pxaq}) has the following structure
\begin{equation}
V(\mathsf{x})  = V_{\mathcal{P}}(\mathsf{x}) + V_{uni}(\mathsf{x}),     \label{V}
\end{equation}
where $V_{\mathcal{P}}(\mathsf{x})$ follows from the form of $a^{-(m-1)(n-1)r}\mathcal{P}_r(a,1,q^{-1})$, while $V_{uni}(\mathsf{x})$ is a universal piece arising from all other contributions in the summand in (\ref{Pxaq})  
\begin{equation}
V_{uni}(\mathsf{x} ) = i \pi  \log \mathsf{x} +f \log^2 \mathsf{x} +2 \log a \log \mathsf{x} -\frac{1}{2} \text{Li}_2\big(-a^{-2} \mathsf{x} ^2 \big)+\frac{1}{2}\text{Li}_2\big(\mathsf{x} ^2\big).  \label{Vuni}
\end{equation}
Also note that the following formulae are useful in determining $V(\mathsf{x})$
\begin{align}
\begin{split}
q^{r^2} & = e^{\frac{1}{\hbar} \log^2 \mathsf{x}}, \qquad a^{r} = e^{\frac{1}{\hbar} \log \mathsf{x} \log a}, \qquad
(\mathsf{x};q^2)_k \sim e^{\frac{1}{2\hbar} ( \text{Li}_2(\mathsf{x}) -  \text{Li}_2(\mathsf{x}z^2) +\ldots  )}  \\
 {r \brack k}_{q^2} & \equiv \frac{(q^2;q^2)_r}{(q^2;q^2)_k(q^2;q^2)_{r-k}} \sim e^{\frac{1}{2\hbar} \big(-\text{Li}_2(\mathsf{x}^2) + \text{Li}_2(z^2)  + \text{Li}_2(\mathsf{x}^2 z^{-2}) +\ldots \big) }
\end{split}
\end{align}
where $\mathsf{x}=q^r$, $z=q^k$, and $\ldots$ denotes terms with higher powers of $\hbar$.

Furthermore, we find that A-polynomials for lattice paths have the following general properties:


\paragraph{Framing change}

To make contact with counting of lattice paths we have to fix framing, which is represented by the factor $q^{fr^2}$ in (\ref{Pxaq}), as $f=mn$. However, we can consider more general classical A-polynomial curves, with arbitrary value of $f$. In this case, changing the framing by $f$ transforms classical A-polynomial $A(x,y,a)$ into 
\begin{equation}
A_{f}(x,y,a)=A(x y^f,y,a).
\end{equation}

\paragraph{Symmetry} 
In addition, there is a symmetry property which roughly switches $x\leftrightarrow x^{-1}$ and $y\leftrightarrow y^{-1}$, in a suitable framing. For the A-polynomials above corresponding to lattice paths below $y=\frac{m}{n}x$, we first decrease framing by $m+n$, 
i.e. we set
\begin{equation}
A'_{m,n}(x,y,a)=A_{m,n}(x y^{-m-n},y,a).
\end{equation}
Then we have
\begin{equation}
A'_{m,n}(x^{-1},-a^2 y^{-1},a) = x^{-\omega} A'_{m,n}(x,y,a),
\end{equation}
for some integer $\omega$. In this case, the number $\omega$ is equal to the number of monomials of the bottom row of the (uncolored) HOMFLY-PT polynomial of the corresponding $(m,n)$ torus knot, i.e.
\begin{equation}
\omega=\frac{1}{m+n} \binom{m+n}{m}.
\end{equation}

The symmetry property is then the statement that $A'_{K}(x^{-1},-a^2 y^{-1},a)$ is equal to $A'_{K}(x,y,a)$ up to multiplication by a monomial.

\paragraph{Specializations}
If we set $a=i$  in $A(x,y,a)$ (i.e. $a^2=-1$ since $a$ appears only with even powers in $A(x,y,a)$), then the obtained polynomial is divisible by $y-1$:
\begin{equation}
A(x,y,a)\mid_{a^2=-1} = (y-1) P(x,y),
\end{equation}
for some polynomial $P(x,y)$ in $x$ and $y$ with integer coefficients. This follows from the fact that setting $a^2=-q$ in the expression for $\bar{P}(x,a,q)$ gives just 1, and therefore in the $q=1$ limit, and consequently $a^2=-1$, we get $y=y(x)=1$, so that $y-1$ must be a factor.\\

In addition if we set $a=i y$  in $A(x,y,a)$ (i.e. $a^2=-y^2$ since $a$ appears only with even powers in $A(x,y,a)$), then the obtained polynomial is also divisible by $y-1$:
\begin{equation}
A(x,y,a)\mid_{a^2=-y^2} = (y-1) Q(x,y),
\end{equation}
for some polynomial $Q(x,y)$ in $x$ and $y$ with integer coefficients. This is a consequence of another canceling differential property of $\mathcal{P}_r$ for the corresponding knot, i.e. from setting $a^2=-q^{2r+1}$. Moreover, for the same framing $f$ as above in the symmetry property, in this specialization $A_f(x,y,a)$ is also divisible by $1+x$. Thus, overall we have
\begin{equation}
A(x,y,a)\mid_{a^2=-y^2} = (1-y)(1+x y^f) R(x,y),
\end{equation}
for some integer $f$ and polynomial $R(x,y)$ in $x$ and $y$ with integral coefficients.  

\bigskip

We present explicit form of classical and quantum A-polynomials for various classes of lattice paths in section \ref{sec-A-results}.


\section{Quivers for generalized Schr\"oder paths}    \label{sec-quivers}

In this section we illustrate our main results stated in propositions \ref{prop1txt} and \ref{prop2txt}. Namely, in several explicit and non-trivial examples, we show that generating functions of generalized Schr{\"o}der paths under a line of rational slope $y=\frac{m}{n}x$ are captured by the same quivers that encode colored invariants of $(m,n)$ torus knots in framing $mn$, as asserted by proposition \ref{prop1txt}. At the same time, these generating functions, by proposition \ref{prop2txt}, follow directly from expressions related to quadruply-graded homology for such knots. 


\subsection{Schr{\"o}der paths for the slope $\frac1f$}    \label{ssec-1f}



To start with, we consider generalized Schr{\"o}der paths under the line $y=x/f$. As the first example, which in fact was the motivation for this work, we recall the relation, found in \cite{Panfil:2018sis}, between Schr\"oder paths (under the diagonal line $y=x$, i.e. for $f=1$) and quiver generating series corresponding to the suitably framed unknot. In this case (and in the conventions and normalization of the current paper)  the full colored HOMFLY-PT polynomials of the unknot in framing $f=1$ are encoded in the following quiver matrix: 
\begin{equation}
C=\left[\begin{array}{cc}1&1\\1&2\end{array}\right]
\end{equation}
for which the quiver series takes form
\begin{equation}P_C(x_1,x_2)=\sum_{i,j}(-1)^{i+j} (-q)^{ i^2+2 ij+2j^2}\frac{x_1^{i}x_2^{j}}{(q^2;q^2)_i(q^2;q^2)_j}.
\end{equation}
Now let  
\begin{equation}\frac{P_C(qx_1,qx_2)}{P_C(q^{-1}x_1,q^{-1}x_2)}=\sum_{i,j}\bar{n}_{i,j}(q)x_1^i x_2^j.
\end{equation}
Further specialization of the from $x_1\to -a^2x$, and $x_2\to x$ yields
\begin{equation}
y(x,a,q) = \frac{P_C(-a^2qx, qx)}{P_C(-a^2q^{-1}x,q^{-1}x)}= \sum_{k\geq 0} N_k(a,q) x^k =  \sum_{i,j,k}n_{i,j,k}a^i q^j x^k.
\end{equation}
As found in \cite{Panfil:2018sis}, and in agreement with proposition \ref{prop1txt}, the numbers $n_{i,j,k}$ count Schr\"oder paths below the diagonal $y=x$, starting from the origin $(0,0)$ and ending at $(k,k)$, with $\frac{i}{2}$ diagonal steps, and with the area between the the line $y=x$ and the path equal to $\frac{j}{2}$. We list $N_k(a,q)$ for $f=1$ in the first section of table \ref{Nkaq-f}. Note that if we keep arbitrary $a$, we get the weighted count of Schr\"oder paths, where the powers of $a$ count the number of diagonal steps. On the other, for $a=0$ we do not allow diagonal steps, so that the counting reduces to Catalan numbers and their $q$-deformation.





 \begin{table}  
\centering
\begin{tabular}{c|c}
\hline
\hline
 Specialization & $N_k(a,q), f=1$ \\
 \hline
--- & $1, a^2+q,a^4+a^2 q^3+2 a^2 q+q^2+q^4,\ldots$ \\
$q=1$ & $1, 1+a^2, 2+3a^2+a^4, \ldots$ \\
$a=1,q=1$ & $1,2,6,22,90,394, \ldots $ \\
$a=0, q=1$ & $1, 1, 2, 5, 14, \ldots $ \\
$a=0$ & $1, q, q^2+q^4, \ldots $ \\
\hline
\hline
 Specialization & $N_k(a,q), f=2$ \\
 \hline
--- & $1, a^2 q+q^2, a^4 q^4+a^4 q^2+a^2 q^7+2 a^2 q^5+2 a^2 q^3+q^8+q^6+q^4,\ldots$ \\
$q=1$ & $1, a^2+1, 2 a^4+5 a^2+3, 5 a^6+21 a^4+28 a^2+12, \ldots$ \\
$a=1,q=1$ & $1,2,10,66,498,4066, \ldots $ \\
$a=0, q=1$ & $1, 1, 3, 12, 55, \ldots $ \\
$a=0$ & $1, q^2, q^8+q^6+q^4, \ldots $ \\
\hline
\hline
 Specialization & $N_k(a,q), f=3$ \\
 \hline
--- & $1, a^2 q^2+q^3, a^4 q^8+a^4 q^6+a^4 q^4+a^2 q^{11}+2 a^2 q^9+2 a^2 q^7+2 a^2 q^5+$\\
 & $+ q^{12}+q^{10}+q^8+q^6,\ldots$ \\
$q=1$ & $1, a^2+1, 3 a^4+7 a^2+4, 12 a^6+45 a^4+55 a^2+22, \ldots$ \\
$a=1,q=1$ & $1,2,14,134,1482,17818, \ldots $ \\
$a=0, q=1$ & $1, 1, 4, 22, 140, \ldots $ \\
$a=0$ & $1, q^3, q^{12}+q^{10}+q^8+q^6, \ldots $ \\
\hline
\hline
 Specialization & $N_k(a,q), f=4$ \\
 \hline
--- & $1, a^2 q^3+q^4,a^4 q^{12}+a^4 q^{10}+a^4 q^8+a^4 q^6+a^2 q^{15}+2 a^2 q^{13}+$ \\
 & $+2 a^2 q^{11}+2 a^2 q^9+2 a^2 q^7+q^{16}+q^{14}+q^{12}+q^{10}+q^8,\ldots$ \\
$q=1$ & $1, a^2+1, 4 a^4+9 a^2+5, 22 a^6+78 a^4+91 a^2+35, \ldots$ \\
$a=1,q=1$ & $1,2,18,226,3298,52450, \ldots $ \\
$a=0, q=1$ & $1, 1, 5, 35, 285, \ldots $ \\
$a=0$ & $1, q^4, q^{16}+q^{14}+q^{12}+q^{10}+q^8, \ldots $ \\
\hline
\hline
\end{tabular}
\caption{Lattice paths numbers $N_k(a,q)$ and their specializations for several values of $k$ and for $f=1,2,3,4$.}.  \label{Nkaq-f}
\end{table}

Now, we find that the generalization of the above case to arbitrary $f$, i.e. to generalized Schr\"oder paths under the line $y=x/f$ (which correspond to unreduced colored HOMFLY-PT polynomial of the $f$-framed unknot) is captured by the following quiver matrix
$$
C=\left[\begin{array}{cc}f&f\\
f&f+1\end{array}\right]
$$
and the corresponding quiver generating series
\begin{equation}
P_C(x_1,x_2)=\sum_{i,j}(-1)^{f(i+j)} (-q)^{f i^2+2f ij+(f+1)j^2}\frac{x_1^{i}x_2^{j}}{(q^2;q^2)_i(q^2;q^2)_j}.
\end{equation}
Taking the specialization $x_1\to -a^2x$, and $x_2\to x$, we get a single variable series
$$\bar{P}_C(x)=P_C(-a^2x,x),$$
and the $q$-weighted paths counts are captured by the following quotient
\begin{equation}\label{for1f}
y(x,a,q) = \frac{\bar{P}_C(qx)}{\bar{P}_C(q^{-1}x)}=\sum_{k\ge 0} N_k(a,q) x^k=\sum_{i,j,k}n_{i,j,k}a^i q^j x^k. 
\end{equation}
With this notation,  we have
\begin{proposition}
The numbers  $n_{i,j,k}$ in (\ref{for1f}) count the generalized Schr\"oder paths under the line $y=\frac{1}{f}x$, starting at $(0,0)$ ending at $(fk,k)$, with $\frac{i}{2}$ diagonal steps, and with the area between the boundary line $y=\frac{1}{f}x$ and the path equal to $\frac{j}{2}$.
\end{proposition}
 
 We assemble explicit form of $N_k(a,q)$ for several values of $f$ and $k$, and also for some specializations of $a$ and $q$, also in Table \ref{Nkaq-f}. Note that the specialization $a=0$ (i.e. ignoring diagonal steps), for a given $f>1$, produces so called $q$-deformed Fuss-Catalan numbers, and setting further $q=1$ yields ordinary Fuss-Catalan numbers.


\subsection{Slope $\frac{2}{3}$}    \label{ssec-23}

Generating series of generalized Schr{\"o}der paths under the line $y=\frac23 x$ are related to the full colored HOMFLY-PT polynomial of torus knot $T_{2,3}$, i.e. trefoil, in framing 6. The quiver corresponding to reduced version of polynomial for the positive 0-framed $T_{2,3}$ torus knot, and the vector $\mathbf{a}$, have been obtained in \cite{Kucharski:2017ogk}: 
$$
C =\left[\begin{array}{ccc}0&1&1\\
1&2&2\\
1&2&3\end{array}\right],\qquad \mathbf{a}=(2,2,4).
$$
It follows that the matrix of the  quiver corresponding to the unreduced HOMFLY-PT polynomial for the negative $T_{2,3}$ in framing $6$ takes form
$$
\bar{C} =\left[\begin{array}{cccccc}6&6&4&5&4&5\\
6&7&4&5&4&5\\
4&4&4&4&3&4\\
5&5&4&5&3&4\\
4&4&3&3&3&3\\
5&5&4&4&3&4
\end{array}\right]
$$
and the corresponding quiver generating series reads
$$
P_{\bar{C}}(x_1,x_2,\ldots,x_6)=\sum_{d_1,d_2,\ldots,d_6} (-1)^{2\cdot 3\cdot \sum_i d_i} (-q)^{\sum_{i,j}\bar{C}_{i,j} d_i d_j}\frac{\prod_i x_i^{d_i}}{\prod_i (q^2;q^2)_{d_i}}.
$$
Upon specializations
$$x_1\to -a^2x,\quad x_2\to x, \quad x_3 \to -a^2x, \quad x_4\to x, \quad x_5 \to a^4x, \quad x_6 \to -a^2x,$$
we get a single variable series
$$\bar{P}_{\bar{C}}(x)=P_{\bar{C}}(-a^2x,x,-a^2x,x,a^4x,-a^2x)$$
and then the quotient 
\begin{equation} \label{for23}  
y(x,a,q) = \frac{\bar{P}_{\bar{C}}(qx)}{\bar{P}_{\bar{C}}(q^{-1}x)}=\sum_{k} N_k(a,q) x^k=\sum_{i,j,k}n_{i,j,k}a^i q^j x^k.
\end{equation}
With this notation, we have
\begin{proposition}
The numbers  $n_{i,j,k}$ in (\ref{for23}) count the generalized Schr\"oder paths under the line $y=\frac{2}{3}x$, starting at $(0,0)$ ending at $(3k,2k)$, with $\frac{i}{2}$ diagonal steps, and with the area between the boundary line $y=\frac{2}{3}x$ and the path equal to $\frac{j}{2}$.
\end{proposition}
 
Explicit form of $N_k(a,q)$ for several values of $k$ is:
\begin{footnotesize}
 $$
 1,a^4 q^2 + 2 a^2 q^3 + q^4 + a^2 q^5 + q^6,a^8 q^4 + 4 a^6 q^5 + 6 a^4 q^6 + a^8 q^6 + 4 a^2 q^7 + 
 6 a^6 q^7 + q^8 + 12 a^4 q^8 + a^8 q^8 + 10 a^2 q^9 + 6 a^6 q^9 + $$
 $$+3 q^{10} + 13 a^4 q^{10} + 12 a^2 q^{11} + 4 a^6 q^{11} + 4 q^{12} + 
 12 a^4 q^{12} + 12 a^2 q^{13} + 2 a^6 q^{13} + 4 q^{14} + 8 a^4 q^{14} + 
 10 a^2 q^{15} + a^6 q^{15} + 4 q^{16} +$$
 $$+ 5 a^4 q^{16} + 7 a^2 q^{17} + 3 q^{18} + 
 2 a^4 q^{18} + 4 a^2 q^{19} + 2 q^{20} + a^4 q^{20} + 2 a^2 q^{21} + q^{22} + 
 a^2 q^{23} + q^{24},a^{12} q^6 + 6 a^{10} q^7 + 15 a^8 q^8 +$$
 $$+  2 a^{12} q^8 + 20 a^6 q^9 + 
 15 a^{10} q^9 + 15 a^4 q^{10} + 45 a^8 q^{10} + 3 a^{12} q^{10} + 6 a^2 q^{11} + 
 70 a^6 q^{11} + 24 a^{10} q^{11} + q^{12} +60 a^4 q^{12} + 78 a^8 q^{12} +$$
 $$+   2 a^{12} q^{12} + 27 a^2 q^{13} + 132 a^6 q^{13} + 25 a^{10} q^{13} + 5 q^{14} + 
 123 a^4 q^{14} + 99 a^8 q^{14} + 2 a^{12} q^{14} + 60 a^2 q^{15} +187 a^6 q^{15} + 24 a^{10} q^{15} +$$
 $$+  12 q^{16} + 187 a^4 q^{16} + 104 a^8 q^{16} + a^{12}q^{16} + 
 96 a^2 q^{17}+ 216 a^6 q^{17} + 20 a^{10} q^{17} + 20 q^{18} + 234 a^4 q^{18}+ 100 a^8 q^{18} + a^{12} q^{18} +$$
 $$+ 128 a^2 q^{19} + 224 a^6 q^{19}+ 16 a^{10} q^{19} + 
 28 q^{20} + 257 a^4 q^{20} + 86 a^8 q^{20} + 148 a^2 q^{21} + 209 a^6 q^{21} + 
 11 a^{10} q^{21} + 34 q^{22} + 256 a^4 q^{22} +$$
 $$+ 72 a^8 q^{22} + 155 a^2 q^{23} + 
 187 a^6 q^{23} + 7 a^{10} q^{23} + 37 q^{24} + 239 a^4 q^{24} + 54 a^8 q^{24}+ 
 150 a^2 q^{25} + 156 a^6 q^{25} +4 a^{10} q^{25} + 37 q^{26} +$$
 $$+  214 a^4 q^{26} + 
 40 a^8 q^{26} + 141 a^2 q^{27} + 126 a^6 q^{27} + 2 a^{10} q^{27} + 36 q^{28} + 
 181 a^4 q^{28} + 26 a^8 q^{28} + 124 a^2 q^{29} +95 a^6 q^{29}+ a^{10} q^{29} + $$
 $$+ 
 33 q^{30} + 149 a^4 q^{30} + 17 a^8 q^{30} + 107 a^2 q^{31} + 69 a^6 q^{31} + 
 29 q^{32} + 116 a^4 q^{32} + 9 a^8 q^{32} + 88 a^2 q^{33} +47 a^6 q^{33} + 25 q^{34} +$$
 $$+  88 a^4 q^{34} + 5 a^8 q^{34} + 71 a^2 q^{35} + 30 a^6 q^{35} + 
 21 q^{36} + 62 a^4 q^{36} + 2 a^8 q^{36} + 54 a^2 q^{37} + 18 a^6 q^{37} + 
 17 q^{38} + 43 a^4 q^{38}+ a^8 q^{38} +$$
 $$+ 40 a^2 q^{39} + 10 a^6 q^{39} + 
 13 q^{40} + 27 a^4 q^{40} + 28 a^2 q^{41} + 5 a^6 q^{41} + 10 q^{42} + 
 17 a^4 q^{42} + 19 a^2 q^{43} + 2 a^6 q^{43}+ 7 q^{44} + 9 a^4 q^{44} +  $$
 $$+
 12 a^2 q^{45} + a^6 q^{45} + 5 q^{46} + 5 a^4 q^{46} + 7 a^2 q^{47}+ 3 q^{48} + 
 2 a^4 q^{48} + 4 a^2 q^{49} + 2 q^{50} + a^4 q^{50} + 2 a^2 q^{51} + q^{52} + 
 a^2 q^{53} + q^{54}, \ldots$$
\end{footnotesize}
For $q=1$ this specializes to
\begin{align}
\begin{split}
&1,2+3a^2+a^4, 23+62a^2+59a^4+23a^6+3a^8, \\
& 377 + 1468 a^2 + 2285 a^4 + 1804 a^6 + 753 a^8 + 155 a^{10} + 12 a^{12}, \ldots
\end{split}
\end{align}
For $a=1$ and $q=1$ we get
$$
1,6,170,6854, \ldots
$$
while for $a=0$ and $q=1$ we get,  as expected, the numbers of Duchon paths
$$
1,2,23,377,7229,\ldots
$$


\subsection{Slope $\frac{2}{5}$}   \label{ssec-25}

Paths under the line $y=\frac25 x$ correspond to the full, unreduced colored HOMFLY-PT polynomials of the $10$-framed negative torus knot $T_{2,5}$ (i.e. $5_1$ knot). The quiver corresponding to reduced version of polynomial for the positive, 0-framed $T_{2,5}$ torus knot, and the vector $\mathbf{a}$, are determined in \cite{Kucharski:2017ogk}:
$$
C =\left[\begin{array}{ccccc}0&1&1&3&3\\
1&2&2&3&3\\
1&2&3&4&4\\
3&3&4&4&4\\
3&3&4&4&5
\end{array}\right],\qquad   \mathbf{a} = (4,4,6,4,6).
$$
It follows that the quiver corresponding to the unreduced HOMFLY-PT polynomial for the negative $T_{2,5}$ in framing $10$ takes form
$$
\bar{C} =\left[\begin{array}{cccccccccc}
10&10&8&9&8&9&6&7&6&7\\
10&11&8&9&8&9&6&7&6&7\\
8&8&8&8&7&8&6&7&6&7\\
9&9&8&9&7&8&6&7&6&7\\
8&8&7&7&7&7&5&6&5&6\\
9&9&8&8&7&8&5&6&5&6\\
6&6&6&6&5&5&6&6&5&6\\
7&7&7&7&6&6&6&7&5&6\\
6&6&6&6&5&5&5&5&5&5\\
7&7&7&7&6&6&6&6&5&6
\end{array}\right]
$$
The corresponding quiver generating series is:
$$
P_{\bar{C}}(x_1,x_2,\ldots,x_{10})=\sum_{d_1,d_2,\ldots,d_{10}} (-1)^{2\cdot 5\cdot \sum_i d_i} (-q)^{\sum_{i,j}\bar{C}_{i,j} d_i d_j}\frac{\prod_i x_i^{d_i}}{\prod_i (q^2;q^2)_{d_i}}.
$$
Upon specializations: 
\begin{align}
\begin{split}
& x_{2i-1} \to -a^2 x_{2i}, \quad i=1,\ldots,5, \\
& x_2\to x,  \quad x_4\to x, \quad x_6 \to -a^2x, \quad x_8 \to x, \quad x_{10} \to -a^2x,
\end{split}
\end{align}
we get a single variable series:
$$\bar{P}_{\bar{C}}(x)=P_{\bar{C}}(-a^2x,x,-a^2x,x,a^4x,-a^2x,-a^2x,x,a^4x,-a^2x)$$
and then the quotient 
\begin{equation}   \label{for25}
y(x,a,q) = \frac{\bar{P}_{\bar{C}}(qx)}{\bar{P}_{\bar{C}}(q^{-1}x)}=\sum_{k} N_k(a,q) x^k=\sum_{i,j,k}n_{i,j,k}a^i q^j x^k.
\end{equation}

\begin{proposition}The numbers  $n_{i,j,k}$ in (\ref{for25}) count the generalized Schr\"oder paths under the line $y=\frac{2}{5}x$, starting at $(0,0)$ ending at $(5k,2k)$, with $\frac{i}{2}$ diagonal steps, and with the area between the boundary line $y=\frac{2}{5}x$ and the path equal to $\frac{j}{2}$.\end{proposition}
 
Explicitly,  coefficients $N_k(a,q)$ for several first values of $k$ read
 \begin{footnotesize}
$$
1,a^4 q^6+a^4 q^4+a^2 q^9+2 a^2 q^7+2 a^2 q^5+q^{10}+q^8+q^6,a^8 q^{24}+a^8 q^{22}+2 a^8 q^{20}+3 a^8 q^{18}+4 a^8 q^{16}+4 a^8 q^{14}+4 a^8 q^{12}+$$
 $$+3 a^8 q^{10}+a^8 q^8+a^6 q^{31}+2 a^6 q^{29}+4 a^6 q^{27}+7 a^6 q^{25}+10 a^6 q^{23}+14 a^6 q^{21}+18 a^6 q^{19}+20 a^6 q^{17}+20 a^6 q^{15}+18 a^6 q^{13}+$$
 $$+12 a^6 q^{11}+4 a^6 q^9+a^4 q^{36}+2 a^4 q^{34}+5 a^4 q^{32}+8 a^4 q^{30}+13 a^4 q^{28}+18 a^4 q^{26}+25 a^4 q^{24}+31 a^4 q^{22}+36 a^4 q^{20}+$$
 $$+37 a^4 q^{18}+36 a^4 q^{16}+30 a^4 q^{14}+18 a^4 q^{12}+6 a^4 q^{10}+a^2 q^{39}+2 a^2 q^{37}+4 a^2 q^{35}+7 a^2 q^{33}+10 a^2 q^{31}+14 a^2 q^{29}+19 a^2 q^{27}+$$
 $$+24 a^2 q^{25}+28 a^2 q^{23}+30 a^2 q^{21}+30 a^2 q^{19}+28 a^2 q^{17}+22 a^2 q^{15}+12 a^2 q^{13}+4 a^2 q^{11}+q^{40}+q^{38}+2 q^{36}+3 q^{34}+4 q^{32}+5 q^{30}+$$
 $$+7 q^{28}+8 q^{26}+9 q^{24}+9 q^{22}+9 q^{20}+8 q^{18}+6 q^{16}+3 q^{14}+q^{12} \ldots
$$
\end{footnotesize}
For $q=1$ this specializes to
\begin{align}
\begin{split}
& 1,2 a^4+5 a^2+3,   23 a^8+130 a^6+266 a^4+235 a^2 + 76, \\ 
& 377 a^{12}+3358 a^{10}+12109 a^8+22715 a^6+23452 a^4+12668 a^2+2803,  
\end{split}
\end{align}
For $a=1$ and $q=1$ we get
\begin{equation}
1,10,730,77482, \ldots
\end{equation}
while for $a=0$ and $q=1$
 \begin{equation} 
1,3,76,2803,121637,\ldots
\end{equation}


    
\subsection{Slope $\frac{3}{4}$}    \label{ssec-34}

Lattice paths under the line $y=\frac34 x$ correspond to the full, unreduced HOMFLY-PT polynomial of the $12$-framed negative torus knot $T_{3,4}$ (i.e. $8_{19}$ knot). The quiver corresponding to reduced version of extremal polynomial (of fize 5, corresponding to 5 generators in the bottom row, and encoding only a part of HOMFLY-PT polynomials that involve extremal powers of $a$) was found in \cite{Kucharski:2017ogk}.  Taking advantage of the expression (\ref{Psuper819}) we determine now the quiver for the positive, 0-framed $T_{3,4}$ torus knot, including the full $a$-dependence of its colored HOMFLY-PT polynomials:  
$$
C=\left[\begin{array}{ccccccccccc}
0&1&2&3&5&1&2&3&4&5&4\\
1&2&3&3&5&2&3&3&5&5&5\\
2&3&4&4&5&3&4&4&5&5&5\\
3&3&4&4&5&4&4&4&6&5&6\\
5&5&5&5&6&6&5&6&6&6&6\\
1&2&3&4&6&3&3&4&5&6&5\\
2&3&4&4&5&3&5&4&6&5&6\\
3&3&4&4&6&4&4&5&6&6&6\\
4&5&5&6&6&5&6&6&7&7&7\\
5&5&5&5&6&6&5&6&7&7&7\\
4&5&5&6&6&5&6&6&7&7&8
\end{array}\right], \qquad  \mathbf{a} = (6,6,6,6,6,8,8,8,8,8,10).
$$
It follows that the matrix $\bar{C}$ of the corresponding quiver for the unreduced HOMFLY-PT polynomial of the negative $T_{3,4}$ in framing $12$ is of size $22\times 22$ and reads:

\arraycolsep 4pt
$$\left[
\begin{array}{cccccccccccccccccccccc}
 12 & 12 & 10 & 11 & 9 & 10 & 8 & 9 & 6 & 7 & 10 & 11 & 9 & 10 & 8 & 9 & 7 & 8 & 6 & 7 & 7 & 8 \\
 12 & 13 & 10 & 11 & 9 & 10 & 8 & 9 & 6 & 7 & 10 & 11 & 9 & 10 & 8 & 9 & 7 & 8 & 6 & 7 & 7 & 8 \\
 10 & 10 & 10 & 10 & 8 & 9 & 8 & 9 & 6 & 7 & 9 & 10 & 8 & 9 & 8 & 9 & 6 & 7 & 6 & 7 & 6 & 7 \\
 11 & 11 & 10 & 11 & 8 & 9 & 8 & 9 & 6 & 7 & 9 & 10 & 8 & 9 & 8 & 9 & 6 & 7 & 6 & 7 & 6 & 7 \\
 9 & 9 & 8 & 8 & 8 & 8 & 7 & 8 & 6 & 7 & 8 & 9 & 7 & 8 & 7 & 8 & 6 & 7 & 6 & 7 & 6 & 7 \\
 10 & 10 & 9 & 9 & 8 & 9 & 7 & 8 & 6 & 7 & 8 & 9 & 7 & 8 & 7 & 8 & 6 & 7 & 6 & 7 & 6 & 7 \\
 8 & 8 & 8 & 8 & 7 & 7 & 8 & 8 & 6 & 7 & 7 & 8 & 7 & 8 & 7 & 8 & 5 & 6 & 6 & 7 & 5 & 6 \\
 9 & 9 & 9 & 9 & 8 & 8 & 8 & 9 & 6 & 7 & 7 & 8 & 7 & 8 & 7 & 8 & 5 & 6 & 6 & 7 & 5 & 6 \\
 6 & 6 & 6 & 6 & 6 & 6 & 6 & 6 & 6 & 6 & 5 & 6 & 6 & 7 & 5 & 6 & 5 & 6 & 5 & 6 & 5 & 6 \\
 7 & 7 & 7 & 7 & 7 & 7 & 7 & 7 & 6 & 7 & 5 & 6 & 6 & 7 & 5 & 6 & 5 & 6 & 5 & 6 & 5 & 6 \\
 10 & 10 & 9 & 9 & 8 & 8 & 7 & 7 & 5 & 5 & 9 & 9 & 8 & 9 & 7 & 8 & 6 & 7 & 5 & 6 & 6 & 7 \\
 11 & 11 & 10 & 10 & 9 & 9 & 8 & 8 & 6 & 6 & 9 & 10 & 8 & 9 & 7 & 8 & 6 & 7 & 5 & 6 & 6 & 7 \\
 9 & 9 & 8 & 8 & 7 & 7 & 7 & 7 & 6 & 6 & 8 & 8 & 7 & 7 & 7 & 8 & 5 & 6 & 6 & 7 & 5 & 6 \\
 10 & 10 & 9 & 9 & 8 & 8 & 8 & 8 & 7 & 7 & 9 & 9 & 7 & 8 & 7 & 8 & 5 & 6 & 6 & 7 & 5 & 6 \\
 8 & 8 & 8 & 8 & 7 & 7 & 7 & 7 & 5 & 5 & 7 & 7 & 7 & 7 & 7 & 7 & 5 & 6 & 5 & 6 & 5 & 6 \\
 9 & 9 & 9 & 9 & 8 & 8 & 8 & 8 & 6 & 6 & 8 & 8 & 8 & 8 & 7 & 8 & 5 & 6 & 5 & 6 & 5 & 6 \\
 7 & 7 & 6 & 6 & 6 & 6 & 5 & 5 & 5 & 5 & 6 & 6 & 5 & 5 & 5 & 5 & 5 & 5 & 4 & 5 & 4 & 5 \\
 8 & 8 & 7 & 7 & 7 & 7 & 6 & 6 & 6 & 6 & 7 & 7 & 6 & 6 & 6 & 6 & 5 & 6 & 4 & 5 & 4 & 5 \\
 6 & 6 & 6 & 6 & 6 & 6 & 6 & 6 & 5 & 5 & 5 & 5 & 6 & 6 & 5 & 5 & 4 & 4 & 5 & 5 & 4 & 5 \\
 7 & 7 & 7 & 7 & 7 & 7 & 7 & 7 & 6 & 6 & 6 & 6 & 7 & 7 & 6 & 6 & 5 & 5 & 5 & 6 & 4 & 5 \\
 7 & 7 & 6 & 6 & 6 & 6 & 5 & 5 & 5 & 5 & 6 & 6 & 5 & 5 & 5 & 5 & 4 & 4 & 4 & 4 & 4 & 4 \\
 8 & 8 & 7 & 7 & 7 & 7 & 6 & 6 & 6 & 6 & 7 & 7 & 6 & 6 & 6 & 6 & 5 & 5 & 5 & 5 & 4 & 5 \\
\end{array}
\right]$$
The corresponding quiver generating series is
$$P_{\bar{C}}(x_1,x_2,\ldots,x_{22})=\sum_{d_1,d_2,\ldots,d_{22}} (-1)^{3\cdot 4\cdot \sum_i d_i} (-q)^{\sum_{i,j}\bar{C}_{i,j} d_i d_j}\frac{\prod_i x_i^{d_i}}{\prod_i (q^2;q^2)_{d_i}}.$$
By taking the specializations: 
\begin{align}
\begin{split} 
& x_{2i-1} \to -a^2 x_{2i}, \qquad i=1,\ldots,11, \\
& x_2,x_4,x_6,x_8,x_{10}\to x,  \quad x_{12},x_{14},x_{16},x_{18},x_{20}\to -a^2x, \quad x_{22} \to a^4x,
\end{split}
\end{align}
we get a single variable series:
\begin{align}
\begin{split}
\bar{P}_{\bar{C}}(x) &= P_{\bar{C}}(-a^2x,x,-a^2x,x,-a^2x,x,-a^2x,x,-a^2x,x,a^4x,-a^2x,a^4x, \\
&\qquad \qquad -a^2x,a^4x,-a^2x,a^4x,-a^2x,a^4x,-a^2x,-a^6x,a^4x),
\end{split}
\end{align}
and then the quotient gives
\begin{equation}   \label{for34}
y(x,a,q) = \frac{\bar{P}_{\bar{C}}(qx)}{\bar{P}_{\bar{C}}(q^{-1}x)}=\sum_{k} N_k(a,q) x^k=\sum_{i,j,k}n_{i,j,k}a^i q^j x^k.
\end{equation} 

\begin{proposition}
The numbers  $n_{i,j,k}$ in (\ref{for34}) count the generalized Schr\"oder paths under the line $y=\frac{3}{4}x$, starting at $(0,0)$ ending at $(4k,3k)$, with $\frac{i}{2}$ diagonal steps, and with the area between the boundary line $y=\frac{3}{4}x$ and the path equal to $\frac{j}{2}$.
\end{proposition}
 
 For arbitrary $a$ and $q$,  $N_k(a,q)$ for a few values of $k$ read
\begin{footnotesize}
 $$1,a^6 q^3+a^4 q^8+2 a^4 q^6+3 a^4 q^4+a^2 q^{11}+2 a^2 q^9+4 a^2 q^7+3 a^2 q^5+q^{12}+q^{10}+2 q^8+q^6,
 a^{12} q^{12}+a^{12} q^{10}+a^{12} q^8+a^{12} q^6+$$
 $$+a^{10} q^{23}+2 a^{10} q^{21}+4 a^{10} q^{19}+6 a^{10} q^{17}+9 a^{10} q^{15}+12 a^{10} q^{13}+12 a^{10} q^{11}+10 a^{10} q^9+6 a^{10} q^7+a^8 q^{32}+2 a^8 q^{30}+5 a^8 q^{28}+$$ $$+9 a^8 q^{26}+16 a^8 q^{24}+24 a^8 q^{22}+35 a^8 q^{20}+44 a^8 q^{18}+53 a^8 q^{16}+55 a^8 q^{14}+49 a^8 q^{12}+35 a^8 q^{10}+15 a^8 q^8+a^6 q^{39}+$$
 $$+2 a^6 q^{37}+5 a^6 q^{35}+10 a^6 q^{33}+18 a^6 q^{31}+29 a^6 q^{29}+45 a^6 q^{27}+63 a^6 q^{25}+84 a^6 q^{23}+105 a^6 q^{21}+120 a^6 q^{19}+$$ $$+127 a^6 q^{17}+120 a^6 q^{15}+96 a^6 q^{13}+60 a^6 q^{11}+20 a^6 q^9+a^4 q^{44}+2 a^4 q^{42}+5 a^4 q^{40}+9 a^4 q^{38}+17 a^4 q^{36}+27 a^4 q^{34}+$$ $$+42 a^4 q^{32}+59 a^4 q^{30}+82 a^4 q^{28}+104 a^4 q^{26}+128 a^4 q^{24}+146 a^4 q^{22}+156 a^4 q^{20}+153 a^4 q^{18}+135 a^4 q^{16}+99 a^4 q^{14}+$$ $$+55 a^4 q^{12}+15 a^4 q^{10}+a^2 q^{47}+2 a^2 q^{45}+4 a^2 q^{43}+7 a^2 q^{41}+12 a^2 q^{39}+19 a^2 q^{37}+28 a^2 q^{35}+38 a^2 q^{33}+51 a^2 q^{31}+65 a^2 q^{29}+$$ $$+78 a^2 q^{27}+90 a^2 q^{25}+97 a^2 q^{23}+98 a^2 q^{21}+92 a^2 q^{19}+76 a^2 q^{17}+52 a^2 q^{15}+26 a^2 q^{13}+6 a^2 q^{11}+q^{48}+q^{46}+2 q^{44}+3 q^{42}+$$ $$+5 q^{40}+7 q^{38}+10 q^{36}+12 q^{34}+16 q^{32}+19 q^{30}+22 q^{28}+24 q^{26}+25 q^{24}+24 q^{22}+22 q^{20}+17 q^{18}+11 q^{16}+5 q^{14}+q^{12},...$$
\end{footnotesize}
Setting $q=1$ we get
\begin{equation}
1, a^6+6 a^4+10 a^2+5, 4 a^{12}+62 a^{10}+343 a^8+905 a^6+1235 a^4+842 a^2+227, \ldots
\end{equation}
For $a=1$ and $q=1$ we get
\begin{equation}
1,22,3618,871510, \ldots 
\end{equation}
while for $a=0$
 \begin{equation}
1,5,227,15090,1182187,\ldots
\end{equation}


\section{A-polynomials for generalized Schr{\"o}der paths}    \label{sec-A-results}

In this section we present explicit results for classical and quantum A-polynomials that encode generating series of generalized Schr{\"o}der paths.  We derive such A-polynomials following general prescriptions presented in section \ref{sec-A}.  

Moreover, anticipating that the relation to combinatorial models generalizes to other knots, and as a prerequisite for future work,  we also derive classical A-polynomials for the generating series (\ref{Pxaq}) involving superpolynomials for $4_1$ and $5_2$ knots. In this case the resulting series $y=y(x)$ also have integral coefficients -- finding combinatorial models that reproduce these results is an interesting challenge.


\subsection{A-polynomial for the slope $\frac{1}{f}$}

Counting of lattice paths under the line of the slope $\frac{1}{f}$ corresponds to the framed unknot.  Generating functions of such lattice paths are encoded by the series (\ref{Pxaq}), in this case simply with $p_r(a,q)=q^{fr^2}$, which represents the framing factor.  Writing (\ref{Pxaq}) as $\bar{P}(x,a,q)=\sum_{r=0}^{\infty} \bar{P}_r x^r$, we have
\begin{equation}
\bar{P}_{r+1} = -q^{(2r+1)f} \frac{a^2 + q^{2r+1}}{1-q^{2r+2}} \bar{P}_r.
\end{equation}
Multiplying both sides by $(1-q^{2r+2})$ and rewriting the resulting expression as a relation for the generating function using operators (\ref{hatx-haty}), produces the quantum A-polynomial
\begin{equation}
\widehat{A}(\hat x,\hat y) = 1 - \hat y + q^f a^2 xy^f + q^{f+1} \hat x \hat y^{f+1}.    \label{qAhat-1f}
\end{equation}
This quantum A-polynomial annihilates (\ref{Pxaq}) with $p_r(a,q)=q^{fr^2}$ (for any chosen framing $f$)  
\begin{equation}
\widehat{A}(\hat x,\hat y) \bar{P}(x,a,q) = 0.
\end{equation}

We can also determine the classical A-polynomial by the saddle point method. To this end we identify $V(\mathsf{x})$ in (\ref{V}) as $V(\mathsf{x}) = V_{uni}(\mathsf{x})$ given in (\ref{Vuni}).  We find that in this case $y=y(x)$, found in section \ref{ssec-1f} and obtained as $q=1$ limit of (\ref{for1f}), or equivalently by proposition \ref{prop2txt}, for any chosen framing $f$, satisfies the A-polynomial equation $A(x,y)=0$ with
\begin{equation}
A(x,y) = 1 - y + a^2xy^f + xy^{f+1}.
\end{equation}
This $A(x,y)$ agrees with $q=1$ limit of (\ref{qAhat-1f}).


\subsection{A-polynomial for the slope $\frac23$}

Lattice paths for the slope $\frac23$ correspond to trefoil in framing $f=6$. To determine quantum A-polynomial for such paths, we consider the series (\ref{Pxaq}) with the contribution (\ref{PQr-trefoil}).  Writing (\ref{Pxaq}) as $\bar{P}(x,a,q)=\sum_r \bar{P}_r x^r$ and using the package \cite{qZeil}, we find a recursion relation for coefficients $\bar{P}_r$
\begin{equation}
\alpha_0 \bar{P}_r + \alpha_1 \bar{P}_{r+1} + \alpha_2 \bar{P}_{r+2} = 0,
\end{equation}
where
\begin{align}
\begin{split}
\alpha_0 & =  q^{18 r+17} \left( q^{2 r+1}+a^2\right) \left(q^{4 r+7} +a^2\right) \\
\alpha_1 & = q^{6 r+3} \left(q^{4 r+5} + a^2\right) \big(a^4 q^6+a^2 q^{2 r+9}+a^2 q^{4 r+9}+a^2 q^{4 r+13}+q^{4 r+10}+ \\
& \qquad   + q^{4 r+12}-q^{6 r+14}+q^{8 r+16}\big) \\
\alpha_2 & = \left(1-q^{2 r+4}\right) \left(q^{4 r+3} + a^2\right) .
\end{split}
\end{align}
Turning this expression into a difference operator acting on the generating function $\bar{P}(x,a,q)$ and writing it in terms of (\ref{hatx-haty}), we find that quantum A-polynomial (\ref{qAhat}) takes form
\begin{equation}
\widehat{A}(\hat x, \hat y) = \beta_0 + \hat{x} \beta_1 + \hat{x}^2 \beta_2,
\end{equation}
where
\begin{align}
\begin{split}
\beta_0 & = 1 - \hat{y} + a^{-2} q^{-5}\hat{y}^2 - a^{-2}q^{-5} \hat{y}^3  \\
\beta_1 & = a^4 q^3 \hat{y}^3 + a^2 q^4 \hat{y}^4 - a^{-2} q^6 \hat{y}^8 + a^{-2} q^6 \hat{y}^9 + q^2 (a^2 + q + a^2q^2 + q^3 + a^2q^4) (\hat{y}^5  + a^{-2} q \hat{y}^7)  \\
\beta_2 & =  a^2 q^{17}\hat{y}^9 + q^{18} \hat{y}^{10} + q^{24} \hat{y}^{11} + a^{-2} q^{25} \hat{y}^{12} .
\end{split}
\end{align}
In the classical limit $q=1$ this expression factorizes
\begin{equation}
\lim_{q\to 1} \widehat{A}(\hat x, \hat y) = (1 + a^{-2}y^2)\big( 1 - y +x  (a^4 y^3+a^2 y^4+2 y^5+2 a^2 y^5-y^6+y^7 ) + x^2  (a^2 y^9+y^{10} )\big).  \label{Axy-23-full}
\end{equation}

\begin{figure}[h]
  \centering
  \includegraphics[width=0.35\textwidth]{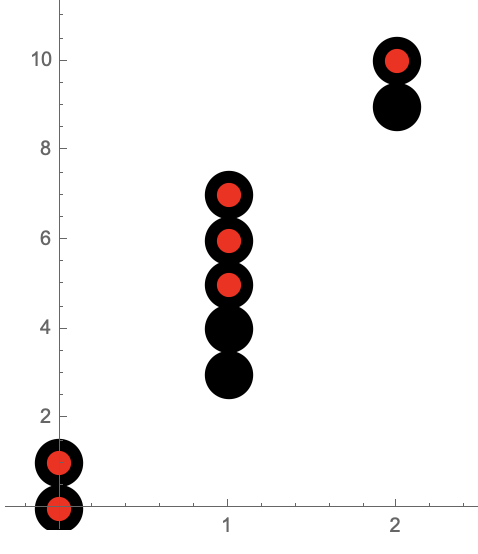}
  \caption{Newton polygons for A-polynomial (\ref{ref-A31}) for lattice paths under the line $y=\frac23 x$.  Black dots represent Newton polygon for generic values of $a$ and red dots represent the case $a=0$.}  \label{fig-newton31}
\end{figure}

We can also determine the classical A-polynomial using the saddle point method, following section (\ref{sec-A}). Again, we consider the series (\ref{Pxaq}) with the contribution (\ref{PQr-trefoil}). Setting $z_1 = q^{k_1}$, we find
\begin{align}
\begin{split}
V_{3_1}(\mathsf{x}) &= V_{uni}(\mathsf{x}) +2 \log ^2 z_1 -4 \log \mathsf{x} \log z_1+ \\
&\quad + \frac{1}{2}\Big( \text{Li}_2\big(z_1^2\big)+\text{Li}_2\big(\mathsf{x}^2z_1^{-2}\big)+ \text{Li}_2\big(-a^2z_1^{-2}\big) -\text{Li}_2\big(\mathsf{x}^2\big) \Big).
\end{split}
\end{align}
It follows that
\begin{equation}
\mathsf{y} = \frac{\mathsf{x}^{2f}(\mathsf{x}^2 + a^2)}{z_1^2 (\mathsf{x}^2 - z_1^2)}, \qquad
1 = \frac{(\mathsf{x}^2-z_1^2)(z_1^2 + a^2)}{\mathsf{x}^4 (z_1^2 - 1)}.
\end{equation}
Eliminating $z_1$ from these equations and removing an irrelevant overall factor according to (\ref{Adual}) we find A-polynomial, which for  $f=6$ takes form
\begin{equation}
A(x,y) = 1 - y +x \big(a^4 y^3+a^2 y^4+2 y^5+2 a^2 y^5-y^6+y^7\big) + x^2 \big(a^2 y^9+y^{10}\big).   \label{ref-A31}
\end{equation}
This is indeed the same as the second factor in (\ref{Axy-23-full}). The equation $A(x,y)=0$ is indeed satisfied by $q=1$ limit of the generating function of lattice paths $y(x,a,q)$ determined in section \ref{ssec-23}. Newton polygons for A-polynomial in  (\ref{ref-A31}) are shown in fig. \ref{fig-newton31}.

\subsection{A-polynomial for the slope $\frac25$}

\begin{figure}[h]
  \centering
  \includegraphics[width=0.35\textwidth]{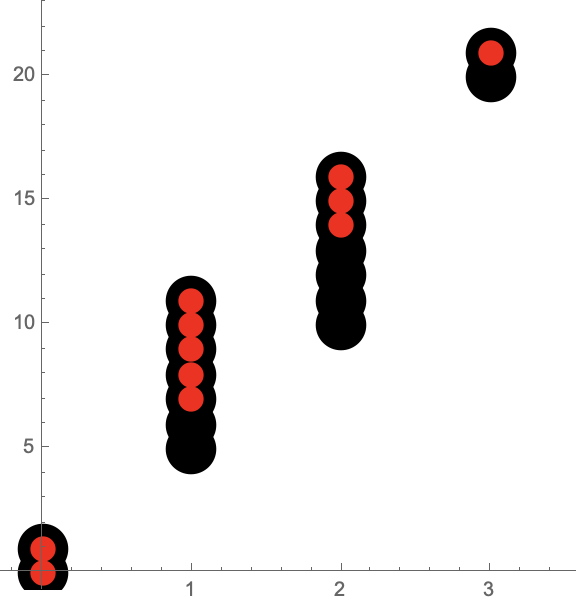}
  \caption{Newton polygons for A-polynomial (\ref{ref-A51}) for lattice paths under the line $y=\frac25 x$.  Black dots represent Newton polygon for generic values of $a$ and red dots represent the case $a=0$.}  \label{fig-newton51}
\end{figure}

In this case we consider just the saddle point method.  For the slope $\frac25$, the generating series (\ref{Pxaq}) involves the contribution (\ref{Pr-22p1}) with $p=2$, and we find that
\begin{equation}
V_{5_1}(\mathsf{x}) = V_{3_1}(\mathsf{x}) + 2\log^2 z_2 -4\log z_2 \log \mathsf{x} -\frac{1}{2}\text{Li}_2(z_1^2) + \frac{1}{2}\text{Li}_2(z_2^2)+ \frac{1}{2}\text{Li}_2(z_1^2z_2^{-2}),
\end{equation}
where $z_1=q^{k_1}, z_2=q^{k_2}$. The system of equations (\ref{saddle-eqs}) takes form
\begin{equation}
\mathsf{y} = \frac{\mathsf{x}^{2f}(\mathsf{x}^2+a^2)}{z_1^2 z_2^4 (\mathsf{x}^2 - z_1^2)},\qquad 1 = \frac{z_2^2(z_1^2+a^2)(\mathsf{x}^2-z_1^2)}{\mathsf{x}^4(z_1^2 - z_2^2)},\qquad 1 = \frac{z_2^2(z_1^2-z_2^2)}{\mathsf{x}^4(z_2^2-1)}.
\end{equation}
To get A-polynomial for paths for the slope $\frac25$ we set $f=2(2p+1)=10$ and find
\begin{align}
\begin{split}
A(x,y) &= 1 - y + x \big(2 a^4 y^5 + y^6(a^2  - a^4)  + y^7(3 + 4 a^2 + a^4) - y^8( 2  + 2 a^2)  + \\
& + y^9(2 + 2 a^2) - y^{10} + y^{11}\big)  + x^2 \big(a^8 y^{10} + a^6 y^{11} + y^{12}(2 a^4 + 2a^6)  +  \\
&+ y^{13}(2 a^2  + 2 a^4) + y^{14}(3 + 4 a^2  + a^4) + y^{15} (a^2 -1)+ 2 y^{16}\big) + x^3 (a^2 y^{20} + y^{21}).  \label{ref-A51}
\end{split}
\end{align}
Newton polygons for this A-polynomial are shown in fig. \ref{fig-newton51}. The above equation is satisfied by the generating function of paths found in section \ref{ssec-25}  
\begin{align}
\begin{split}
y &= 1 + x (2 a^4 + 5 a^2 + 3) + x^2 (23 a^8 + 130 a^6 + 266 a^4 + 235 a^2 + 76) + \\
 &+ x^3 (377 a^{12} + 3358 a^{10} + 12109 a^8 + 22715 a^6 + 23452 a^4 + 12668 a^2 + 2803) + \ldots
\end{split}
\end{align}

\subsection{A-polynomial for the slope $\frac27$}

In this case we also consider just the saddle point method. For the slope $\frac27$, the generating series (\ref{Pxaq}) involves the contribution (\ref{Pr-22p1}) with $p=3$, and we find that
\begin{equation}
V_{7_1}(\mathsf{x}) = V_{5_1}(\mathsf{x}) + 2\log^2 z_3 - 4\log z_3 \log \mathsf{x} 
-\frac{1}{2}\text{Li}_2(z_2^2) + \frac{1}{2}\text{Li}_2(z_3^2)+ \frac{1}{2}\text{Li}_2(z_2^2z_3^{-2}).
\end{equation}
The generating function of paths under the line of the slope $\frac27$ arises once we set $f=14$ in the term $V_{5_1}(\mathsf{x})$. We then find
\begin{align}
\begin{split}
A(x,y) &= 1 - y + x \Big(3 a^4 y^7 + a^2 y^8 - 2 a^4 y^8 + 6 a^2 y^9 + 2 a^4 y^9 - 
    4 a^2 y^{10} - a^4 y^{10} + 4 a^2 y^{11} +  \\
&+ a^4 y^{11} - 2 a^2 y^{12} + 2 a^2 y^{13} + y^9 (4 + (y-1) y (3 + 2 y^2 + y^4))\Big) + \\
&+ x^2 \Big(3 a^8 y^{14} + 2 a^6 y^{15} - a^8 y^{15} + 6 a^4 y^{16} + 8 a^6 y^{16} + 
    2 a^8 y^{16} + 2 a^4 y^{17} - a^6 y^{17}  + \\
&+ 10 a^4 y^{18} + 4 a^6 y^{18} +  a^4 y^{19} + 2 a^4 y^{20} + 
    a^2 y^{17} (3 + y (12 + y (-2 + y (8 + y)))) + \\
&+    y^{18} (6 + y (-3 + y (6 + y (-2 + 3 y))))\Big) + \\
&+ x^3 \Big(a^{12} y^{21} + 3 a^4 y^{25} + 4 a^4 y^{26} + 2 a^4 y^{27} + 
    a^{10} y^{22} (1 + 2 y) + a^8 y^{23} (2 + y (2 + y)) + \\
&+    a^2 y^{26} (3 + 2 y (3 + y)) + a^6 y^{24} (2 + y (4 + y)) + 
    y^{27} (4 + y (-1 + 3 y))\Big) + \\
&+ x^4 (a^2 y^{35} + y^{36}).   \label{ref-A71}
\end{split}
\end{align}
The above equation is satisfied by the generating function of paths 
\begin{equation}
y = 1 + x (4 + 7 a^2 + 3 a^4) +\ldots
\end{equation}

\begin{figure}[h]
  \centering
  \includegraphics[width=0.5\textwidth]{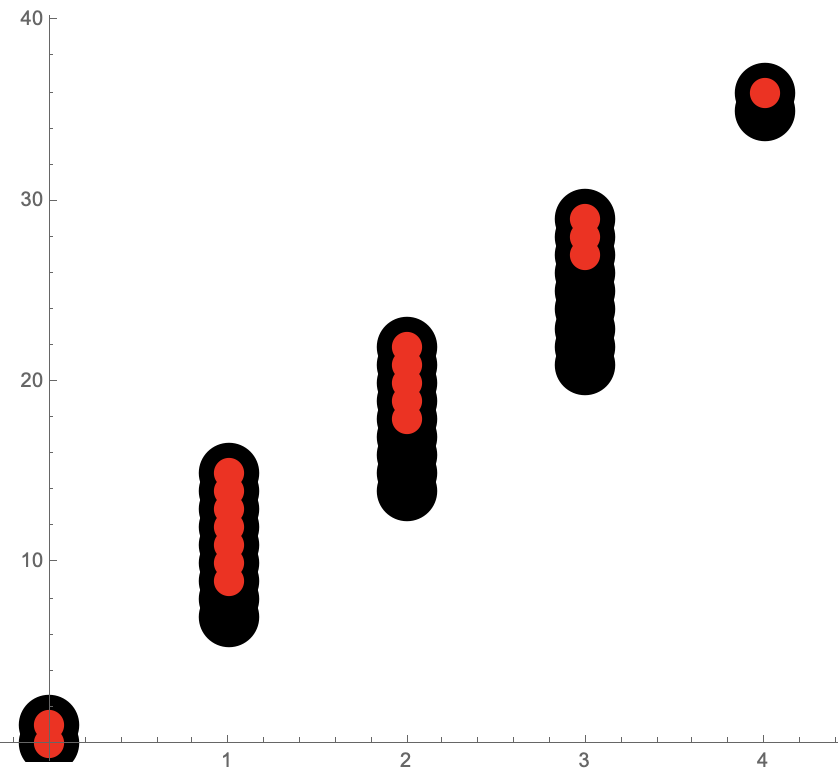}
  \caption{Newton polygons for A-polynomial (\ref{ref-A71}) for lattice paths under the line $y=\frac27 x$.  Black dots represent Newton polygon for generic values of $a$ and red dots represent the case $a=0$.}  \label{fig-newton71}
\end{figure}

Newton polygons for this A-polynomial are shown in fig. \ref{fig-newton71}.  In fact, from figures \ref{fig-newton31}, \ref{fig-newton51} and \ref{fig-newton71} we can deduce the general pattern how Newton polygons look like for A-polynomials for lattice paths under the lines $y=\frac{2}{2p+1}x$, for all $p=0,1,2,\ldots$. Namely, involving full $a$-dependence, i.e. regarding Newton polygons with black dots,  they consist of leftmost and rightmost columns of length 2, and $p$ intermediate columns of length $2p+1$, all at linearly increasing heights.

\subsection{A-polynomial for the slope $\frac34$}

For the lattice paths under the line $y=\frac34 x$ (corresponding to torus knot $T_{3,4}$, i.e. $8_{19}$ knot), we also determine just the classical A-polynomial by the saddle point method. The potential (\ref{V}) takes form
\begin{align}
\begin{split}
V_{8_{19}}(\mathsf{x}) & = V_{uni}(\mathsf{x}) - 6\log^2 \mathsf{x} +2 ( \log^2z_a + \log^2 z_b + \log^2 z_c  ) + \\
&+ 2(\log z_a + \log z_b + \log z_c)(\log z_j - \log z_c) + 2(\log z_j - \log z_c)^2 + \\
&+\frac{1}{2}\big( \text{Li}_2(\mathsf{x}^2 z_j^{-2})  - \text{Li}_2(\mathsf{x}^2) + \text{Li}_2(z_j^2 z_c^{-2}) + \text{Li}_2(z_c^2 z_b^{-2}) + \text{Li}_2(z_a^2) + \text{Li}_2(z_b^2 z_a^{-2})   \big) + \\
&+\frac{1}{2}\big( -\text{Li}_2(-a^2\mathsf{x}^{-2})  + \text{Li}_2(-a^2\mathsf{x}^{-2}z_j^{-2}) + \text{Li}_2(-a^2 z_c^2 z_j^{-2}) \big),
\end{split}
\end{align}
where $z_j=q^j, z_a=q^a,z_b=q^b,z_c=q^c$. The generating function of lattice paths under the line of the slope $\frac34$ arises for $f=12$. We find that A-polynomial takes form
\begin{align}
\begin{split}
A(x,y) & = 1 - y + x \big(a^6 y^4 + a^4 y^5 + a^2 y^6 + 2 a^4 y^6 + 5 y^7 + 9 a^2 y^7 + 3 a^4 y^7 + \\
& - 4 y^8 - 4 a^2 y^8 + y^9 + a^2 y^9 + 3 y^{10} + 3 a^2 y^{10} - y^{12} + y^{13}\big) + \\
& +  x^2 \big(-a^8 y^{10} - a^6 y^{11} + 3 a^4 y^{12} + 4 a^6 y^{12} + 4 a^2 y^{13} + 8 a^4 y^{13} + 3 a^6 y^{13} +\\
& + 27 a^2 y^{14} + 17 a^4 y^{14} -    10 a^2 y^{15} - 4 a^4 y^{15} + 5 a^2 y^{16} + 3 a^4 y^{16} + 6 a^2 y^{17} + \\
& +      y^{14} (10 + y (-6 + y (3 + y (5 + (-1 + y) y))))\big) + \\
& + x^3 \big(-a^{10} y^{16} - a^8 y^{17} - 5 a^6 y^{18} - 6 a^8 y^{18} + 3 a^4 y^{19} + 5 a^6 y^{19} + 3 a^8 y^{19} + \\
&+ 6 a^2 y^{20} + 10 a^4 y^{20} + 4 a^6 y^{20} + 27 a^2 y^{21} + 17 a^4 y^{21} - 8 a^2 y^{22} - 3 a^4 y^{22} + \\
&+ 4 a^2 y^{23} + y^{21} (10 + y (-4 + y (3 + y - y^2)))\big) +  \\
& + x^4 \big(a^{12} y^{22} + a^{10} y^{23} - 3 a^8 y^{25} + a^6 (-3 + y) y^{25} +  4 a^2 y^{27} + 9 a^2 y^{28} - 2 a^2 y^{29} + \\
& +   a^4 y^{26} (1 + y) (1 + 3 y) - y^{28} (-5 + y - y^2 + y^3)\big) + x^5 (a^2 y^{34} + y^{35}).    \label{ref-A819}
\end{split}
\end{align}
The equation $A(x,y)=0$ is solved by the generating function of paths found in section \ref{ssec-34}
\begin{align}
\begin{split}
y(x) &= 1 + (a^6 + 6 a^4 + 10 a^2 + 5)x + (227 + 842 a^2 + 1235 a^4 + 905 a^6 + 343 a^8 + \\
&+ 62 a^{10} + 4 a^{12}) x^2  + (15090 + 81812 a^2 + 190339 a^4 + 248089 a^6 + 198317 a^8 +\\
&+ 99975 a^{10} + 31434 a^{12} + 5857 a^{14} + 575 a^{16} + 22 a^{18}) x^3 + \ldots
\end{split}
\end{align}

\begin{figure}[h]
  \centering
  \includegraphics[width=0.5\textwidth]{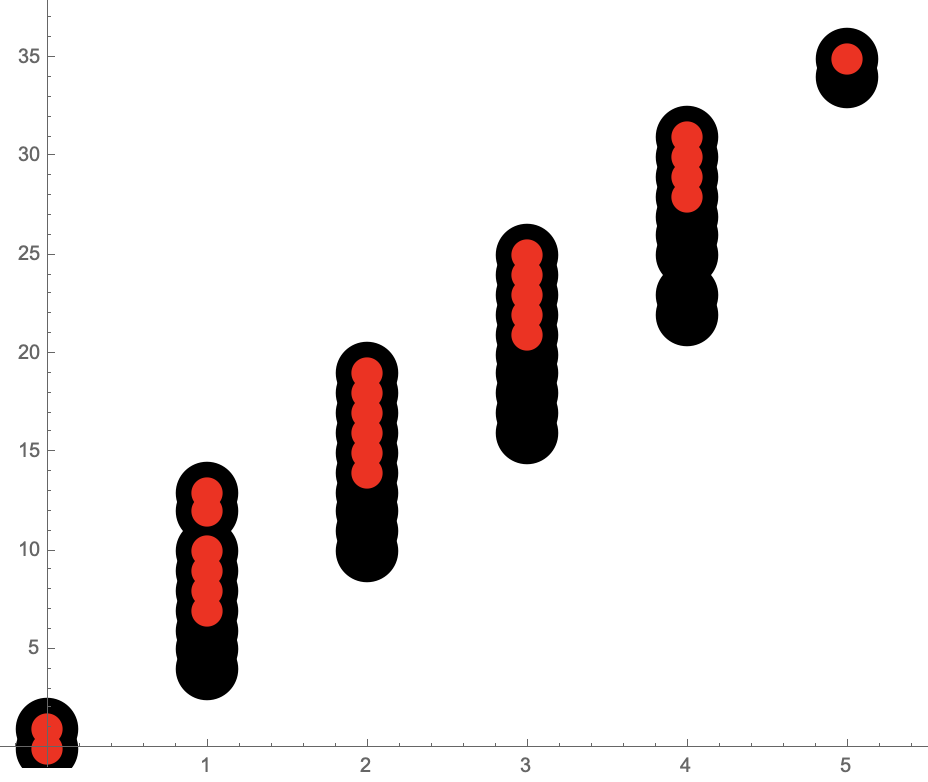}
  \caption{Newton polygons for A-polynomial (\ref{ref-A819}) for lattice paths under the line $y=\frac34 x$.  Black dots represent Newton polygon for generic values of $a$ and red dots represent the case $a=0$.}  \label{fig-newton819}
\end{figure}

Newton polygons for this A-polynomial are shown in fig. \ref{fig-newton819}.  Note that, regarding generic $a$-dependence (i.e.  the polygon with black dots),  there is an interesting symmetry: in the second and the fifth column (counting from the left), the third dot (respectively from the top or bottom) is missing.  Such a symmetry is not manifest for $a=0$ (i.e. for the polygon made of red dots) -- this indicates that $a$-dependent path counts and corresponding A-polynomials are indeed more fundamental.

\subsection{$4_1$ knot}

An interesting challenge is to generalize the relation to lattice paths (or possibly more general combinatorial models) beyond torus knots.  As the first non-trivial example we find A-polynomial curve corresponding to $4_1$ knot; we stress this is not the knot theory A-polynomial (found already e.g. in \cite{superA}), but A-polynomial that arises from the parameter identification introduced in section \ref{ssec-quadruply}, which is relevant for combinatorial models discussed in this paper.  Using (\ref{P-41}) and analogous specializations as in (\ref{Pxaq}), and keeping $f$ arbitrary, the potential (\ref{V}) takes form
\begin{align}
\begin{split}
V_{4_1}(\mathsf{x}) & =  V_{uni}(\mathsf{x})-2\log\mathsf{x}\log z + 2\log a \log z + \frac{1}{2}\big(  -\text{Li}_2(\mathsf{x}^{2})  +  \text{Li}_2(z^2) + \\
& \quad +  \text{Li}_2(\mathsf{x}^2 z^{-2})-\text{Li}_2(-a^{-2}z^2)  + \text{Li}_2(-a^{-2}\mathsf{x}^2) - \text{Li}_2(-a^{-2}\mathsf{x}^2 z^{2})   \big),
\end{split}
\end{align}
where $z=q^k$. In this case the system of equations (\ref{saddle-eqs}) takes form 
\begin{equation}
y = \frac{x^{2f} (\mathsf{x}^2z^2 + a^2) }{\mathsf{x}^2 - z^2},\qquad 1 = \frac{(\mathsf{x}^2 - z^2)(z^2+a^2)(\mathsf{x}^2z^2+a^2)}{a^2\mathsf{x}^2 z^2 (z^2-1)}.
\end{equation}
We then find
\begin{align}
\begin{split}
A(x,y) & = a^2 y^2 (1- y) + x (a^6 y^f + 4 a^4 y^{2 + f} + 4 a^2 y^{3 + f} + y^{5 + f}) + 
 x^2 (-a^6 y^{2 f} + \\
& \quad  - 4 a^4 y^{2 + 2 f} + 4 a^4 y^{3 + 2 f}  +  a^2 y^{5 + 2 f}) + x^3 (-a^6 y^{2 + 3 f} - a^4 y^{3 + 3 f}).
\end{split}
\end{align}
The equation $A(x,y)=0$ is satisfied by
\begin{align}
\begin{split}
y(x)&= 1 + (4 + a^{-2} + 4 a^2 + a^4) x + (1 + a^2) (3 + a^{10} (-2 + f) + f + \\
&\quad +    a^8 (-6 + 7 f) + a^2 (13 + 7 f) + a^6 (1 + 17 f) + a^4 (16 + 17 f)) a^{-4} x^2 + \ldots
\end{split}
\end{align}
It would be great to find a combinatorial model,  in which counting of some objects would reproduce (possibly for some specific $f$) integer coefficients in the above $y(x)$.


\subsection{$5_2$ knot}

Analogously we find A-polynomial curve for $5_2$ knot (again we stress that it is different from the knot theory A-polynomial determined in \cite{FGSS}). In this case we consider the superpolynomial (\ref{P-52}), and the specializations as in (\ref{Pxaq}) yield the potential (\ref{V}) 
\begin{align}
\begin{split}
V_{5_2}(\mathsf{x}) & =  V_{uni}(\mathsf{x})+ 2\log a \log z_j + 2\log a \log z_k - \log^2 z_k -2\log^2 z_j-2\log\mathsf{x}\log z_k  + \\
& + \frac{1}{2}\big( -\text{Li}_2(-a^{-2}z_k^2)   +  \text{Li}_2(-a^{-2}\mathsf{x}^{2})  -  \text{Li}_2(-a^{-2}\mathsf{x}^2 z_j^2)  \big)  + \\
& + \frac{1}{2}\big( \text{Li}_2(\mathsf{x}^{-2})   -  \text{Li}_2(z_k^{-2})  -  \text{Li}_2(z_k^2\mathsf{x}^{-2}) + \text{Li}_2(\mathsf{z_k}^{-2})   -  \text{Li}_2(z_j^{-2})  -  \text{Li}_2(z_j^2 z_k^{-2})  \big)  
\end{split}
\end{align}
where $z_j=q^j,z_k=q^k$. In this case the system of equations (\ref{saddle-eqs}) takes form 
\begin{equation}
y = \frac{x^{2f} (\mathsf{x}^2z_j^2 + a^2) }{z_k^2(\mathsf{x}^2 - z_k^2)},\qquad 1 = \frac{(\mathsf{x}^2 - z_k^2)(z_k^2+a^2)}{\mathsf{x}^4 (z_k^2-z_j^2)},\qquad 1 = \frac{(\mathsf{x}^2z_j^2 + a^2) (z_k^2-z_j^2) }{z_j^2z_k^2(z_j^2-1)}.
\end{equation}
We then find for $f=0$ 
\begin{align}
\begin{split}
A(x,y) & = y^7(1-y)  +  x \big(a^6 y^2 - a^2 y^4 + 4 a^4 y^4 + 2 a^2 y^5 + 2 y^6 + 
    5 a^2 y^6 + (1 - y) y^6 + \\
&+ (y-1) y^7 + y^8\big) + x^2 \big(-a^8 - 3 a^2 y^3 + 4 a^2 y^4 + y^5 + a^2 y^5 - 2 (y-1) y^5 + \\
&+    y^6 - 4 a^2 y^6 + (1 - y) y^7 - a^6 y (1 + 4 y) - a^4 y^2 (1 + y + 6 y^2)\big) + \\
& + x^3 \big(-2 a^6 - 3 a^2 y^2 + 2 a^2 y^3 -  4 a^2 y^4 + (1 - y) y^4 + (1 - y) y^5 - a^4 y (2 + 5 y)\big) +  \\
& + x^4 (-a^4 - a^2 y) .
\end{split}
\end{align}
Interestingly, for $a=0$ this expression factorizes
\begin{equation}
A(x,y)\big|_{a=0} = y^4 (x  - xy + y)  \big(x^2 + x (2 + x) y + y^2 + (x-1) y^3\big).
\end{equation}
Finding a combinatorial model in which the numbers of certain states would be captured by the above A-polynomials is an important challenge too.



\section*{Acknowledgements}

The work of P.S. has been supported by the OPUS grant no.  2022/47/B/ST2/03313 ``Quantum
geometry and BPS states'' funded by the National Science Centre, Poland. The work of M.S. has been supported by the Science Fund of the Republic of Serbia, Project no. 7749891, GWORDS -- ``Graphical Languages", as well as by Funda\c{c}\~ao para a Ci\^encia e a Tecnologia (FCT) through CEEC grant with DOI no. 10.54499/2020.02453.CEECIND/CP1587/CT0007.


\newpage

\bibliographystyle{JHEP}

\bibliography{abmodel}

\end{document}